\documentclass[preprint]{aastex63}

\usepackage{natbib}
\usepackage{color}
\usepackage{graphicx}

\newcommand{\Msun}{M_\odot}

\submitjournal{ApJ}

\begin{document}
	\title{The Post-impact Evolution of the X-ray Emitting Gas in SNR 1987A Viewed by XMM-Newton}
	
	\correspondingauthor{Lei Sun}
	\email{l.sun@uva.nl}
	
	\author{Lei Sun}
			\affiliation{Department of Astronomy, Nanjing University, Nanjing 210023, China}
			\affiliation{Anton Pannekoek Institute, GRAPPA, University of Amsterdam, PO Box 94249, 1090 GE Amsterdam, The Netherlands}
	
	\author{Jacco Vink}
	\affiliation{Anton Pannekoek Institute, GRAPPA, University of Amsterdam, PO Box 94249, 1090 GE Amsterdam, The Netherlands}
	\affiliation{SRON, Netherlands Institute for Space Research, Sorbonnelaan 2, 3584 CA Utrecht, The Netherlands}
	
	\author{Yang Chen}
			\affiliation{Department of Astronomy, Nanjing University, Nanjing 210023, China}
			\affiliation{Key Laboratory of Modern Astronomy and Astrophysics, Nanjing University, Ministry of Education, China}
	
	\author{Ping Zhou}
	\affiliation{Department of Astronomy, Nanjing University, Nanjing 210023, China}
	\affiliation{Anton Pannekoek Institute, GRAPPA, University of Amsterdam, PO Box 94249, 1090 GE Amsterdam, The Netherlands}
	
	\author{Dmitry Prokhorov}
	\affiliation{Anton Pannekoek Institute, GRAPPA, University of Amsterdam, PO Box 94249, 1090 GE Amsterdam, The Netherlands}
	\affiliation{School of Physics, University of the Witwatersrand, 1 Jan Smuts Avenue, Braamfontein, Johannesburg, 2050 South Africa}
	
	\author{Gerd P{\"u}hlhofer}
	\affiliation{Institut f{\"u}r Astronomie und Astrophysik, Eberhard Karls Universit{\"a}t
		T{\"u}bingen, Sand 1, D-72076 T{\"u}bingen, Germany}
	
	\author{Denys Malyshev}
	\affiliation{Institut f{\"u}r Astronomie und Astrophysik, Eberhard Karls Universit{\"a}t
		T{\"u}bingen, Sand 1, D-72076 T{\"u}bingen, Germany}
	
	\begin{abstract}
	Since 1996 the blast wave driven by SN 1987A has been interacting with the dense circumstellar material, which provides us with a unique opportunity to study the early evolution of a newborn supernova remnant (SNR). Based on the XMM-Newton RGS and EPIC-pn X-ray observations from 2007 to 2019, we investigated the post-impact evolution of the X-ray emitting gas in SNR 1987A. The hot plasma is represented by two non-equilibrium ionization components with temperature of $\sim0.6$\,keV and $\sim2.5$\,keV. The low-temperature plasma has a density $\sim2400$\,cm$^{-3}$, which is likely dominated by the lower density gas inside the equatorial ring (ER). The high-temperature plasma with a density $\sim550$\,cm$^{-3}$ could be dominated by the \ion{H}{2} region and the high-latitude material beyond the ring. In the last few years, the emission measure of the low-temperature plasma has been decreasing, indicating that the blast wave has left the main ER. But the blast wave is still propagating into the high-latitude gas, resulting in the steadily increase of the high-temperature emission measure. In the meantime, the average abundances of N, O, Ne, and Mg are found to be declining, which may reflect the different chemical compositions between two plasma components. We also detected the Fe K lines in most of the observations, showing increasing flux and centroid energy. We interpret the Fe K lines as from a third hot component, which may come from the reflected shock-heated gas or originate from Fe-rich ejecta clumps, shocked by the  reverse shock.
	\end{abstract}
	\keywords{ISM: supernova remnants --- supernovae: general --- X rays: individual (SN 1987A) --- X-rays: ISM}
	
	\section{Introduction}
	As the {nearest} supernova (SN) observed since Kepler's SN of 1604, SN 1987A provides us with a unique opportunity to {observe in detail the onset of supernova remnant (SNR) formation and subsequent evolution for {an SN} whose observational properties are known in detail.}
	{SN1987A is located in the Large Magellanic Cloud (LMC) and detected on February 23, 1987.}
	The progenitor star was identified as a blue supergiant (Sanduleak $-69^{\circ}~202$). SN 1987A is surrounded by a peculiar circumstellar material (CSM) system represented by three coaxial rings, the origin of which may be explained by the merging of a binary system at about 20,000 yr before the explosion \citep{1992ApJ...391..246P,2007Sci...315.1103M} {or by 
		collisions of a fast wind colliding with material from an earlier slow wind, during the progenitor's post main sequence evolution \citep{2008AA...488L..37C}}. Today, over 30 years after the SN outburst, the blast wave has encountered and heavily interacted with the CSM structures, especially the dense equatorial ring (ER). The impact of the blast wave on the ER resulted in a complex shock system, which compressed and heated the gas, {giving} rise to radiation in different energy bands. As a result, SN 1987A has made its transition to the remnant phase \citep[see][for a recent review]{2016ARAA..54...19M}. 
	
	As a newborn remnant, SNR 1987A has during its short life already gone through several dynamics phases,
	as monitored regularly across the electromagnetic spectrum.
	{Observations in the X-ray band} are {in particular} 
	crucial for understanding the underlying physics of the shock {interaction} {processes}, as it contains most of the emitting energy.
	The X-ray emission of SNR 1987A was first detected by ROSAT $\sim1500$ days after the explosion and it steadily brightened until $\sim3000$ days, as the blast wave began interacting with the \ion{H}{2} region interior to the ER  \citep{1996AA...312L...9H}. Starting from $\sim4000$ days, the blast wave encountered the dense clumps protruding from the inner edge of the ER, as evidenced by the appearance of several ``hot spots'' in {the} optical band \citep[e.g.,][]{1998ApJ...492L.139S,2000ApJ...537L.123L}. Corresponding to this ---as first observed by Chandra since $\sim4600$ days--- the soft X-ray flux of SNR 1987A was found to 
	exceed the linear extrapolation of the ROSAT light curve and 
	{brighten}
	more rapidly than observed by ROSAT \citep{2000ApJ...543L.149B,2002ApJ...567..314P}. 
	{At $\sim6000$ days the blast wave impacted the main body of the ER, as }
	indicated by a dramatic increase in the soft X-ray flux \citep[e.g.,][]{2005ApJ...634L..73P} and a sudden decrease in the shock velocity \citep[e.g.,][]{2009ApJ...703.1752R}. Thereafter, the soft X-ray flux continued increasing ---first exponentially and then linearly \citep{2013ApJ...764...11H}--- until day $\sim9500$, 
	{when the light curve started to level off}
	\citep{2016ApJ...829...40F}. Together with the optical/infrared light curves, this indicates that the blast wave 
	{had traversed through the entire} dense ring \citep{2015ApJ...806L..19F,2016AJ....151...62A}.

	{Recently,} the overall X-ray light curve has  been well reproduced by three-dimensional hydrodynamic/magnetohydrodynamic simulations \citep[e.g.,][]{2015ApJ...810..168O,2019AA...622A..73O,2020AA...636A..22O}. 
	{These simulations indicate that the X-ray emission is dominated} {first} by the shocked \ion{H}{2} region, then by {the} shocked dense ring, and that {SNR 1987A} will enter a third phase around 32--34 years after the explosion, which 
	{will be}
	dominated by the SN ejecta heated by the reverse shock. 
	{According to these simulations},
	we are expecting to 
	find the X-ray features from the {shocked} ejecta in the most recent observations, or in the near future. On the other hand, recent optical observations revealed new features of spot-like and diffuse emission outside the ER (since $\sim9500$ days), which may be related to high-latitude material that extends from the ER toward the outer rings \citep{2019ApJ...886..147L}. Therefore, we are also expecting to observe the associated X-ray emission from this material. In addition, a recent (2016--2018) enhancement of the GeV emission from the SNR 1987A region has been reported based on Fermi/LAT observations \citep{2019arXiv190303045M}. The X-ray observations could be crucial to clarify the nature of the GeV signal \citep[a detection of the non-thermal emission in the 10--20 keV energy band has been recently reported by][which may be indicative of a pulsar wind nebula]{2021arXiv210109029G}.
	
	In this work, {we report our analysis of} XMM-Newton observations of SNR 1987A taken {during} the last decade, aiming at the post-impact evolution of the X-ray emitting gas. We performed detailed spectral modeling and constrained multiple physical parameters to unveil the plasma properties. 
	{The recent changes in the plasma properties clearly show that the blast wave has left the ER and is propagating into the high-latitude gas.}
	This paper is organized as follows: Section \ref{sec:obs} describes the observations and data reduction procedure, Section \ref{sec:spec} presents the spectral modeling and the main results, {Section \ref{sec:disc}} discusses the further implications of the results, and {Section \ref{sec:con}} gives a brief summary.
	
	\section{Observations and Data Reduction}{\label{sec:obs}}
	
	SNR 1987A has been regularly monitored by XMM-Newton from 2007 to 2017 (PI: F.~Haberl). Recently, we obtained a new observation in November 2019 (PI: Malyshev, for the H.E.S.S. collaboration), aiming at its latest evolution. Thereby, we got a set of data containing 12 observations in total, covering a time interval over 12 years, which is summarized in Table \ref{tab:obs}. {In this work, we utilized these data to analyze the X-ray spectra of SNR 1987A taken in the last decade. The high-resolution RGS spectra provide tight constraints to the plasma properties. We also included the EPIC-pn spectra for a complete energy coverage to 10\,keV following \citet{2010AA...515A...5S}.}
	
\begin{figure*}
		\plotone{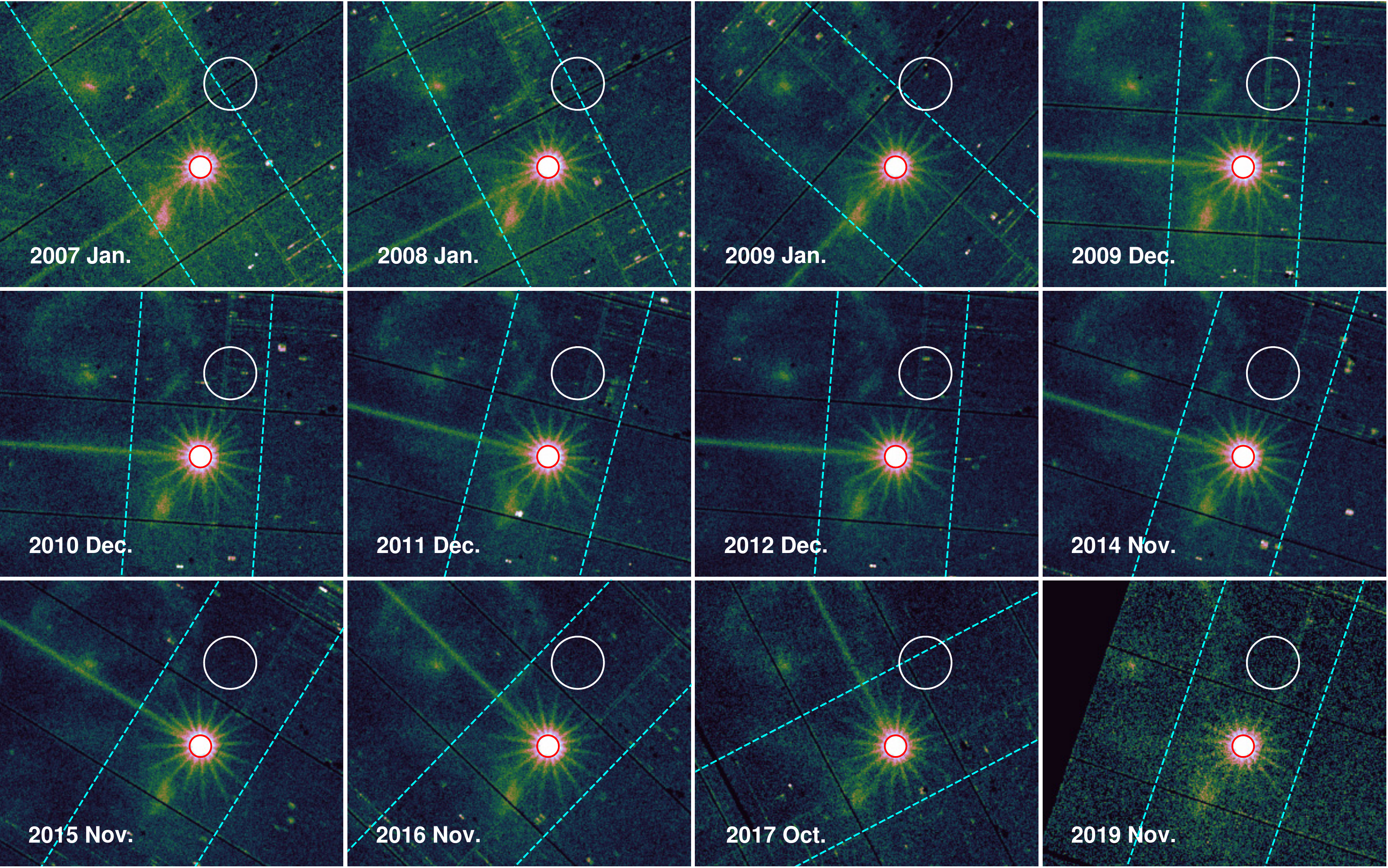}
		\caption{EPIC-pn images of SNR 1987A region. The source and background extraction regions used for the spectral analysis are denoted by red and white circles, respectively. The cyan dashed lines show the RGS dispersion direction and cross-dispersion aperture. \label{fig:region}}
	\end{figure*}

	We used XMM-Newton Science Analysis Software (SAS, version 16.1.0)\footnote{https://www.cosmos.esa.int/web/xmm-newton/sas} to process the data. For EPIC-pn data, the task {\tt epchain} was used to produce calibrated photon event files, and then {\tt pn-filter} was used to filter out soft proton (SP) flares and to remove affected time intervals. The EPIC-pn spectra of SNR 1987A were extracted from a circular region centered on the source with a radius of $25''$ \footnote{SNR 1987A is a point-like source as viewed by EPIC-pn. However, in view of the PSF ($FWHM\sim12.5''$), a circular region with a radius of $25''$ can get most of the source photons involved while avoid large background contamination.}, while the background spectra were extracted from a source-free region nearby (shown in Figure \ref{fig:region}). {We note that the background contributes only $\sim1\%$ to the EPIC-pn spectrum of SNR 1987A, so the background subtraction is only minor contribution to the systematic error budget.}
	We selected only the single events (${\tt PATTERN}=0$) for the source spectra in order to get rid of the photon pile-up effect. The RGS data were processed using {\tt rgsproc}, and then the spectra were extracted within the low-background time intervals (${\tt RATE}<0.2$). For a reliable application of the $\chi^2$ statistics, both of the EPIC-pn and RGS spectra were rebinned using FTOOL\footnote{https://heasarc.gsfc.nasa.gov/ftools/} task {\tt grppha} to obtain at least 25 counts per bin. 
	
	\section{Spectral analysis}\label{sec:spec}
	
	In order to better trace the evolution of the X-ray properties of the remnant, we used a consistent method to analyze all the observations. We used XSPEC (version 12.10.1)\footnote{https://heasarc.gsfc.nasa.gov/xanadu/xspec/} with AtomDB 3.0.9\footnote{http://www.atomdb.org/} for spectral analysis. Unless otherwise specified, in this paper we present metal abundances with respect to their solar values \citep[][hereafter W00]{2000ApJ...542..914W}.
	
	\subsection{Individual emission lines and their fluxes}\label{sec:line}
	
\begin{figure*}
		\plotone{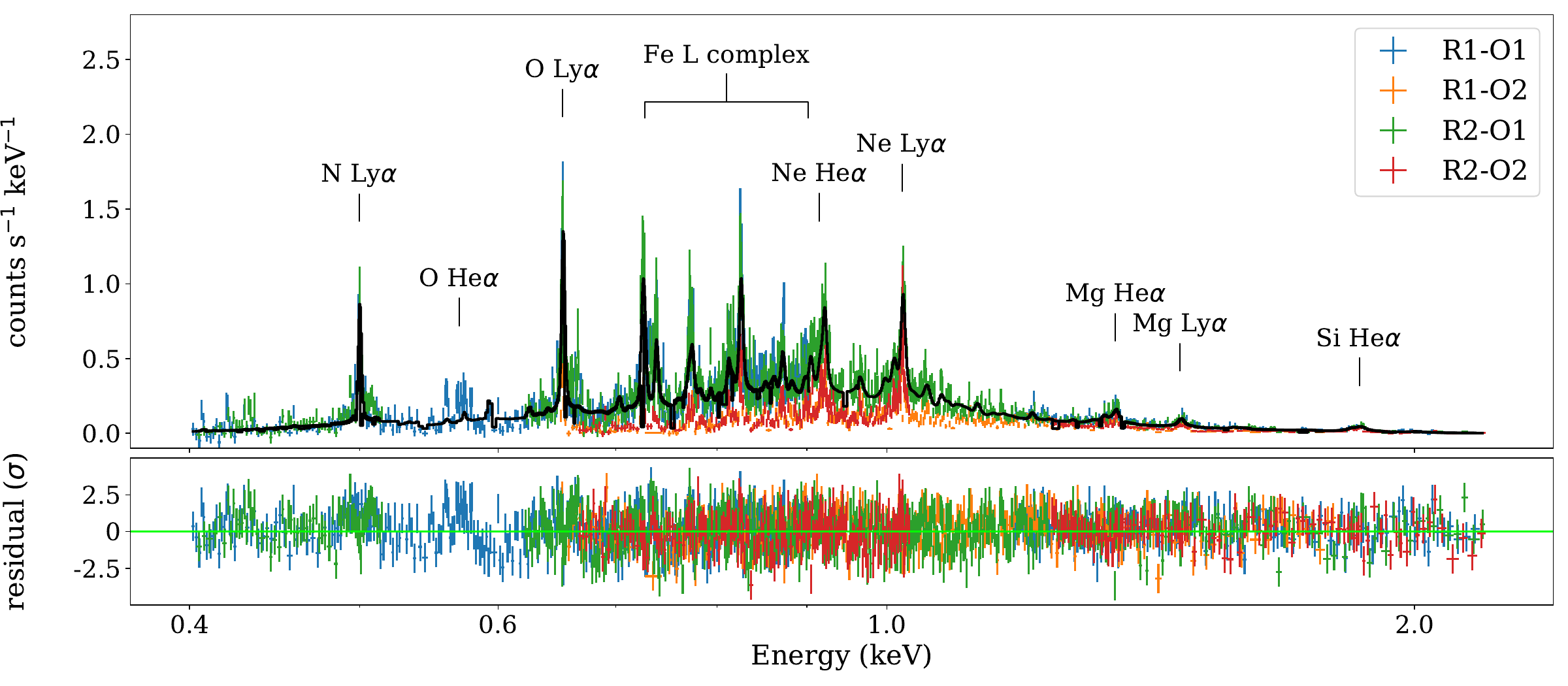}
		\caption{An example of the RGS spectra of SNR 1987A (2015), with bright emission lines labeled. Observed spectra are denoted by data points, with different colors standing for different instruments and grating orders. The best-fit model {({\tt nlapec} $+$ {\tt gauss})} is denoted by the black line, and the residuals are showed in the bottom panel.\label{fig:RGS_spec}}
	\end{figure*}

	Taking advantage of the high spectral resolution of RGS, we {are} 
	able to identify the individual emission lines and to constrain their fluxes. The $0.35$--$2.5$\,keV RGS spectra of SNR 1987A are dominated by the emission lines, which mainly come from the $\alpha$-elements such {as} O, Ne, Mg and Si, and 
	from the Fe-L {transitions} around $0.7$--$0.9$\,keV (Figure \ref{fig:RGS_spec}). In order to identify these lines, we constructed an {heuristic}
	model and fit it to the first-order and the second-order RGS1/RGS2 spectra simultaneously.
	
	The spectral model {consists} of a thermal continuum component (described by the {\tt nlapec} model in XSPEC, including thermal bremsstrahlung, radiative recombination {continua} and two photon emission), and several Gaussian line profiles ({\tt gauss}). Both the {continuum} and the emission lines were subjected to foreground absorption ({\tt tbabs}), for which the Galactic absorption was fixed at $N_{\rm H,\,Gal}=6\times10^{20}$\,cm$^{-2}$ with abundances set to be solar, and the LMC absorption was fixed at $N_{\rm H,\,LMC}=2.2\times10^{21}$\,cm$^{-2}$ with abundances set to be the average LMC abundances given by \citet{1992ApJ...384..508R}. Previous studies have revealed line broadening features in the dispersed X-ray spectra of SNR 1987A \citep[e.g.,][]{2005ApJ...628L.127Z,2019NatAs...3..236M}. Despite its complex origins, the line broadening can be approximately described {with} a {power-law} function of energy \citep[e.g.,][]{2010AA...515A...5S,2012ApJ...752..103D}. Therefore, we simply adopted a Gaussian smoothing component ({\tt gsmooth}) for the broadening effect. The {\tt gsmooth} function convolves the spectrum with a Gaussian function, of which the Gaussian sigma is energy dependent: $\sigma=\sigma_6(E/6\,{\rm keV})^\alpha$, where $\sigma_6$ is the width at 6\,keV. We also introduced a redshift component ({\tt vashift}) to account for the line shift {which may result from the systemic motion of the remnant.} 
	
\begin{figure*}
		\plotone{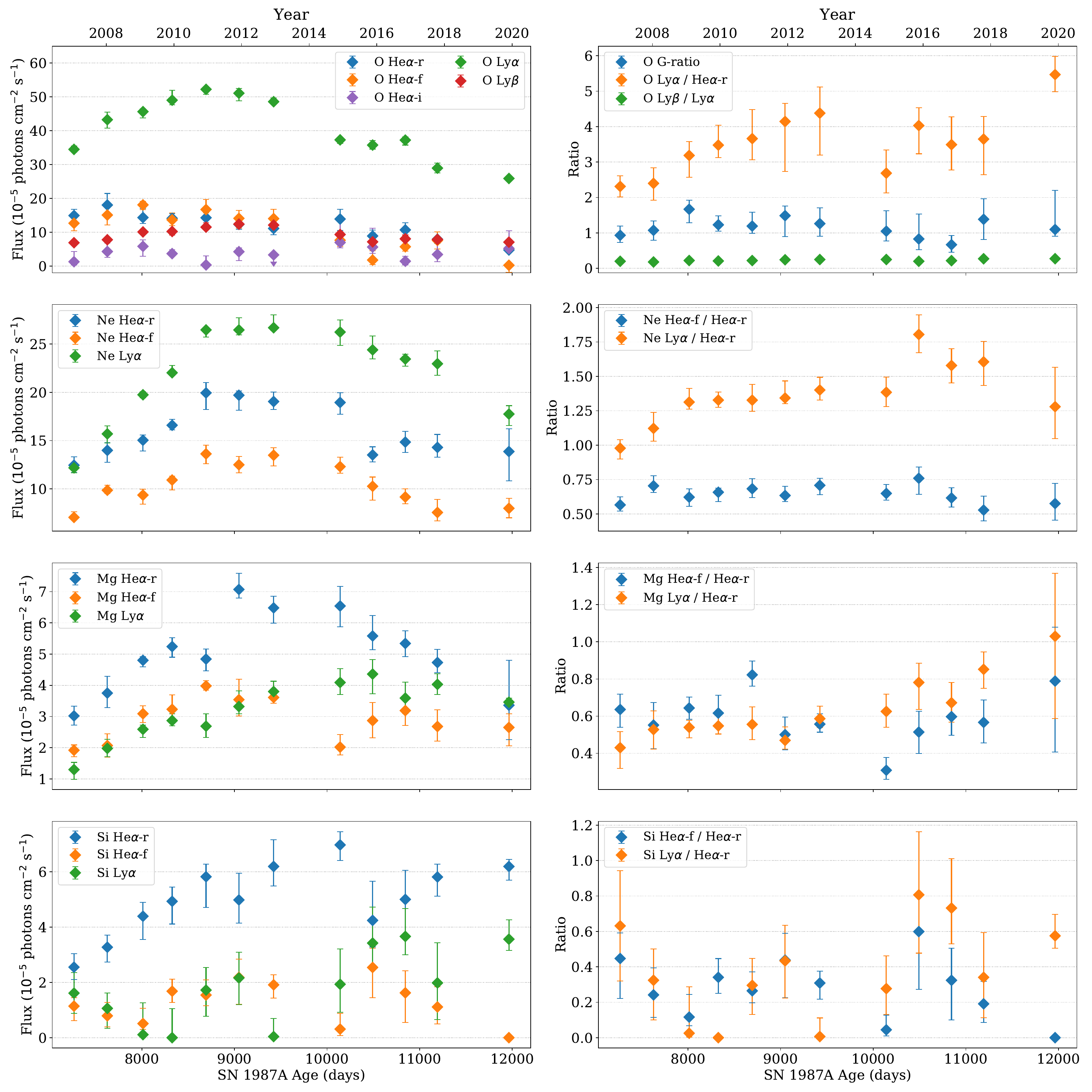}
		\caption{The left panels show the light curves of major emission lines from O, Ne, Mg, and Si. The right panels show the line ratios. Error bars represent the 1-$\sigma$ uncertainties.\label{fig:line_ratio}}
	\end{figure*}

	With {this heuristic}
	spectral model described above, we identified 36 emission lines in the RGS spectra of SNR 1987A. The emission lines, together with their rest frame centroids and (absorption-corrected) fluxes, are listed in Table \ref{tab:line}. Among these emission lines, the He-like triplets and the H-like Ly lines are especially useful {for} thermal X-ray plasma diagnostics. In Figure \ref{fig:line_ratio}, we plotted the light curves of various emission lines from {the} {He- and H-like} ions of O, Ne, Mg, and Si. Most of the emission lines follow a similar {flux-}evolution pattern, {showing initially a brightening},
	{reaching}
	a peak value at around 2011-2015, and {after that a drop-off in flux}. 
	This is consistent with the overall soft X-ray light curve of SNR 1987A \citep[e.g.,][see also Section \ref{sec:lc} below for our updates {with} the new data]{2016ApJ...829...40F}. For the oxygen lines, we calculated the G-ratios (defined as $(f+i)/r$, where $f$, $i$, and $r$ stand for the {fluxes} of the forbidden line, the intercombination lines and the resonance line, respectively), the Ly$\alpha$/He$\alpha$-r flux ratios, and the Ly$\beta$/Ly$\alpha$ flux ratios. For Ne, Mg, and Si, we calculated the He$\alpha$-f/He$\alpha$-r flux ratios and the Ly$\alpha$/He$\alpha$-r flux ratios. As shown in Figure \ref{fig:line_ratio}, we find that while the G-ratios (or He$\alpha$-f/He$\alpha$-r) show no significant variation during this {12-year} interval, the Ly$\alpha$/He$\alpha$ flux ratios have been continuously increasing. This {could result from} the increase in the average plasma temperature, but {may also be} affected by non-equilibrium ionization (NEI) effects.

	{Our spectral analysis also provides measurements of the line broadening and the systemic velocity, see $\sigma_6$ and $v_{\rm 87A}$ in Table \ref{tab:line}.}
	Except for {an} obviously divergent {value} from the 2019 observation (which has the shortest effective exposure and thus the lowest photon statistics), the {line shift} {indicates} a systemic velocity of SNR 1987A {around} 250\,km\,s$^{-1}$, {consistent} with the value derived from optical observations \citep[$286.74\pm0.05$\,km\,s$^{-1}$, e.g.,][]{2008AA...492..481G}. 
	{We note that the systematic error could be significantly greater than the statistical error in this case for systemic velocity measurement. The RGS wavelength scale accuracy is 5\,m\AA\ and 4\,m\AA\ in the first and second order, respectively. Since most of the bright emission lines lie in the range of 10--20\,\AA, a $\sim5$\,m\AA\ uncertainty in the line position will transfer to $\sim70$--$150$\,km\,s$^{-1}$ uncertainty in the systemic velocity. Taking this systematic error into consideration, we estimated the overall uncertainty in systemic velocity as $\sim100$\,km\,s$^{-1}$, which may explain the rather large dispersion of the measured $v_{\rm 87A}$ values.}
	{{On the other hand,} the line shift could also be affected by the asymmetric distribution and expansion of the X-ray emitting gas. As revealed by Chandra observations \citep[][Figure 5 therein]{2016ApJ...829...40F}, the surface brightness distribution of the ER has been dynamically changing since the impact, which may {also} lead to the large dispersion of the values we obtained here for $v_{\rm 87A}$.} The line broadening effect {is} characterized by a Gaussian smooth function, by which we found the {power-law} index $\alpha\sim2$, {similar to the value obtained by \citet{2010AA...515A...5S}, and consistent with the Chandra LETG results \citep{2009ApJ...692.1190Z}.}
	{The power-law index provides us further information about the origin of the line broadening \citep[see also the discussion in][]{2012ApJ...752..103D}. The source extent will result in a broadening which is constant in wavelength and thus gives $\alpha=2$ in energy. This is nearly the case for LETG/HETG, but not for RGS since SNR 1987A is spatially unresolved by XMM-Newton. On the other hand, the Doppler broadening (including the bulk Doppler broadening and the thermal broadening) gives $\alpha=1$ assuming a constant velocity over all the spectral lines. Our result of $\alpha\sim2$ ($>1$) indicates that the spectral lines at higher energy (which are from the heavier elements) have larger velocities. This is consistent with the results obtained by \citet{2019NatAs...3..236M}, in the sense of a proportional relation between the post-shock ion temperature and the ion mass.}
	\subsection{Global spectral fitting}
	
	In order to better characterize the physical properties of the X-ray emitting gas in SNR 1987A, we 
	{fitted} the RGS (0.35--2.5\,keV) and EPIC-pn (0.3--10.0\,keV) spectra simultaneously with detailed {NEI} plasma emission models. 
	
	\subsubsection{two-temperature {\tt vnei} model}\label{sec:vnei}
	{While the actual distribution of temperatures, densities and other plasma properties can be quite complex,}
	previous studies {\citep[e.g.,][]{2010AA...515A...5S,2016ApJ...829...40F}} have shown that the X-ray emission of SNR 1987A can be {well approximated} by a spectral model containing two plasma components: 
	a low-temperature component associated with the dense ring material, and a high-temperature component associated with the lower density gas. Therefore, we {first fit} the RGS and pn spectra with a two-temperature NEI plasma model, described by two {\tt vnei} models in XSPEC. The {\tt vnei} model characterizes the average thermal and ionization state of the plasma by a single electron temperature $kT_{\rm e}$ and a single ionization parameter $\tau=n_{\rm e}t$, respectively, {where $n_{\rm e}$ is the electron density and $t$ is the elapsed time since the gas was shocked}. Following \citet[][and the references therein, {with} abundances converted to W00]{2010AA...515A...5S}, the {abundances} of N, O, Ne, Mg, Si, S, and Fe were set as free parameters, while those of He (set to 2.57), C (0.14), Ar (0.76), Ca (0.49), and Ni (0.98) were fixed, {since they produce no significant features in our data}. The abundances of the two components were assumed to be the same. {The Galactic absorption and the LMC absorption were fixed at $N_{\rm H,\,Gal}=6\times10^{20}$\,cm$^{-2}$ and $N_{\rm H,\,LMC}=2.2\times10^{21}$\,cm$^{-2}$, respectively\footnote{As previously pointed out by \citet{2012ApJ...752..103D}, there could be a measurable absorption component local to SNR 1987A: a density of $1\times10^4$\,H\,cm$^{-3}$ and a path length of 0.01\,pc would result in a local absorption $N_{\rm H,\,local}\sim0.3\times10^{21}$\,cm$^{-2}$. This local absorption is also affected by clumping, the ionization state, and the geometry of the remnant, and thus may change as the system evolves. By setting the LMC absorption as a free parameter among different epochs, we obtain $N_{\rm H,\,LMC}$ varying in the range of $1.7$--$2.5\times10^{21}$\,cm$^{-2}$, with a rather monotonic increasing trend. This may reflect the local absorption, but will introduce additional parameter degeneracy, especially between $N_{\rm H}$, $kT_{\rm e}$, and the normalization. On the other hand, the changes and the calibration uncertainties in the RGS effective area may also affect the measure of $N_{\rm H}$. As a result, we decide to keep $N_{\rm H,\,LMC}$ fixed to its average value ($2.2\times10^{21}$\,cm$^{-2}$) in our spectral fitting, and actually this will not significantly affect our conclusions.}.} Similar to the model described in Section \ref{sec:line}, the spectra were convolved with {\tt gsmooth} function and shifted by {\tt vashift} to account for {the} line broadening and systemic velocity, {respectively}. The parameters of {\tt gsmooth} and {\tt vashift} were fixed at the {values} obtained above by {the} RGS spectral fitting (Table \ref{tab:line}) for each observation.
	
	This two-temperature {\tt vnei} model {gives}
	acceptable fits to the spectra for all the observation, {with reduced chi-square {values of} $\chi^2_{\rm r}\sim1.21$--$1.41$ (dof $\sim1234$--$2433$).} The low-temperature component {is characterised by} $kT_{\rm e,LT}\sim0.6$\,keV and the high-temperature component has $kT_{\rm e,HT}\sim2.5$\,keV, which show no significant variations during the last decade. The detailed fitting results are summarized in Table \ref{tab:2-T nei_nH}. 
	
	\subsubsection{Two-temperature {\tt vpshock} model}\label{sec:vpshock}
	In addition to the {\tt vnei} model presented above, we also {modeled the global spectra of SNR 1987A spectra with }
	a plane-parallel shock model \citep[][described by the {\tt vpshock} model in XSPEC]{2001ApJ...548..820B}. 
	This model describes the X-ray emission from the hot gas heated by a plane-parallel shock with a constant post-shock temperature. It differs from the {\tt vnei} model in its linear distribution of the ionization parameter $\tau$ versus emission measure, {instead of assuming a single ionisation parameter.}
	{The NEI plasma in the {\tt vpshock} model is}
	characterized by a lower limit $\tau_{\rm l}=n_{\rm e}t_{\rm l}$ at the shock front and an upper limit $\tau_{\rm u}=n_{\rm e}t_{\rm u}$ for the earliest shocked gas. We fixed $\tau_{\rm l}$ at zero {and let} $\tau_{\rm u}$ vary freely. Other parameters were set {in} the same way as those in the {\tt vnei} model.
	
	We {find} that the two-temperature {\tt vpshock} model fits the data better, which is indicated by an average drop of the reduced chi-square $\Delta\chi^2_{\rm r}\sim0.07$. This suggested that the single ionization state scenario in {\tt vnei} model is over-simplified, while the plane-parallel shock model
	better represents the actual conditions. However, the best-fit values of the parameters show no significant {discrepancy} from those in the {\tt vnei} fitting, except for {the values of }$\tau_{\rm u}$, which are typically one order of magnitude higher than the average $\tau$ given by the {\tt vnei} fitting (Table \ref{tab:2-T pshock_nH}). {We further discuss the physical implications of the difference between $\tau$ and $\tau_{\rm u}$ in Section \ref{sec:ave_pro}}

	\subsection{X-ray light curve}\label{sec:lc}
	The X-ray light curves of SNR 1987A up to 2015 {have} been reported by \citet{2016ApJ...829...40F} based on both Chandra and XMM-Newton observations. Here we update the recent changes in X-ray fluxes {with} the new XMM-Newton data.
	
	Based on our two-temperature {\tt vpshock} fitting described above, we measured the 0.5--2.0\,keV and 3.0--8.0\,keV fluxes of the remnant from 2015 to 2019, and {show} them in Figure \ref{fig:lc} together with the earlier results. {As noticed by} \citet{2016ApJ...829...40F}, the soft band flux of SNR 1987A reached a plateau at around 9000 days, and {has} remained approximately constant since then. The new observations show that after around 10500 days, the soft band flux started to {decline}. In the latest observation (11964 days), the soft band flux was $\sim6.4\times10^{-12}$\,erg\,cm$^{-2}$\,s$^{-1}$, a drop of 
	$\sim18\%$ {from} its peak value ($\sim7.8\times10^{-12}$\,erg\,cm$^{-2}$\,s$^{-1}$).
	{In the meantime,} the hard band flux kept increasing over the last few years, {with a latest value of $\sim1.5\times10^{-12}$\,erg\,cm$^{-2}$\,s$^{-1}$}. As a result, the hard to soft flux ratio continued climbing up as before (see Figure \ref{fig:lc}).
	
	\subsection{Fe K lines}\label{sec:Fe_K}
\begin{figure*}
		\plotone{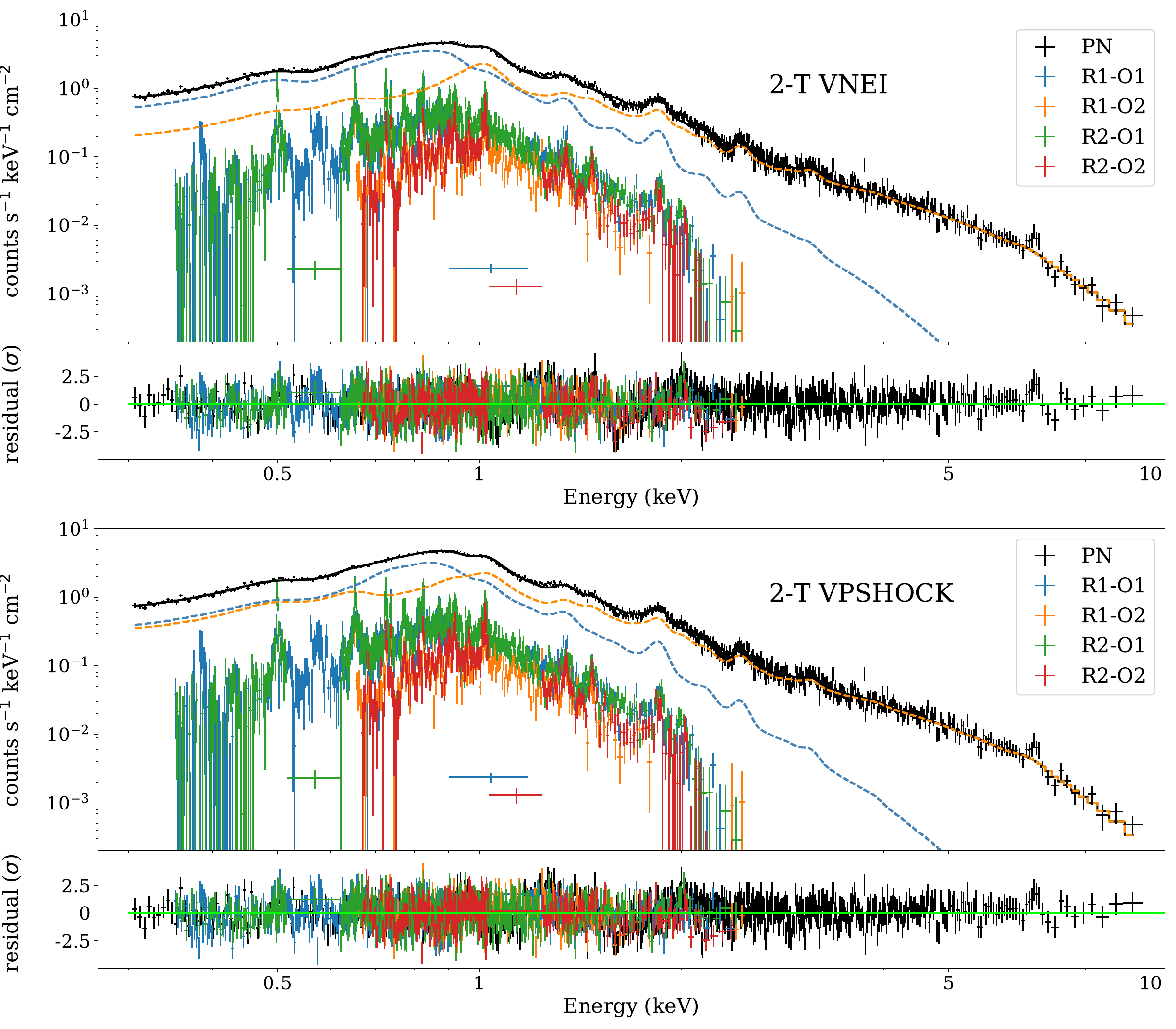}
		\caption{An example of the RGS and EPIC-pn spectra of SNR 1987A (2015), with different spectral models and residuals. Observed spectra are denoted by data points, with different colors standing for different instruments and grating orders. The best-fit models are denoted by black lines, and the high- and low-temperature components by orange and blue dashed lines, respectively. Top: the two-temperature {\tt vnei} model fitting. Bottom: the two-temperature {\tt vpshock} model fitting. \label{fig:spec_fit}}
	\end{figure*}

\begin{figure*}
		\plotone{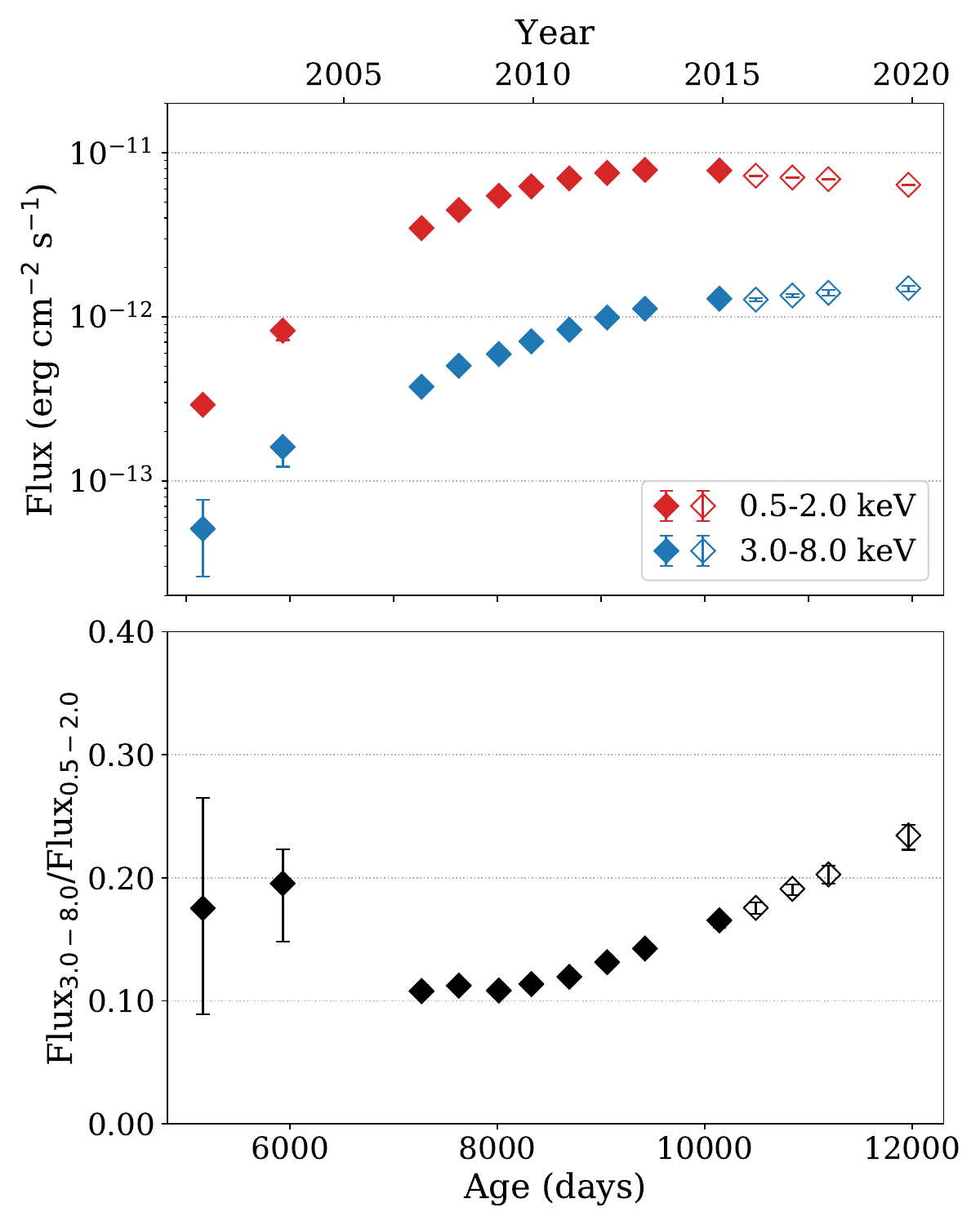}
		\caption{Top: the X-ray light curve of SNR 1987A viewed by XMM-Newton. The 0.5--2.0\,keV fluxes are in blue and the 3.0--8.0\,keV fluxes are in red. Bottom: the hard band to soft band flux ratios. The solid points are obtained from \citet{2016ApJ...829...40F}, while the blank ones are newly published data obtained in this work. Error bars represent the 90\% uncertainties, note that for most of the data points the error bars are too small to be visible. \label{fig:lc}}
	\end{figure*}

	We note that although the global spectra of SNR 1987A can be approximately characterized {with} a two-temperature {\tt vnei} or {\tt vpshock} model, both of these two models {leave}
	some residuals around $\sim6.6$\,keV for the EPIC-pn spectra, corresponding to Fe K line emission (see Figure \ref{fig:spec_fit}). The detection of the Fe K lines has been reported by several studies \citep[e.g.,][]{2010AA...515A...5S,2012AA...548L...3M}, {but} its origin remains unclear. Here, we present a comprehensive investigation {of} the Fe K lines based on the XMM-Newton observations.
	
\begin{figure*}
		\plotone{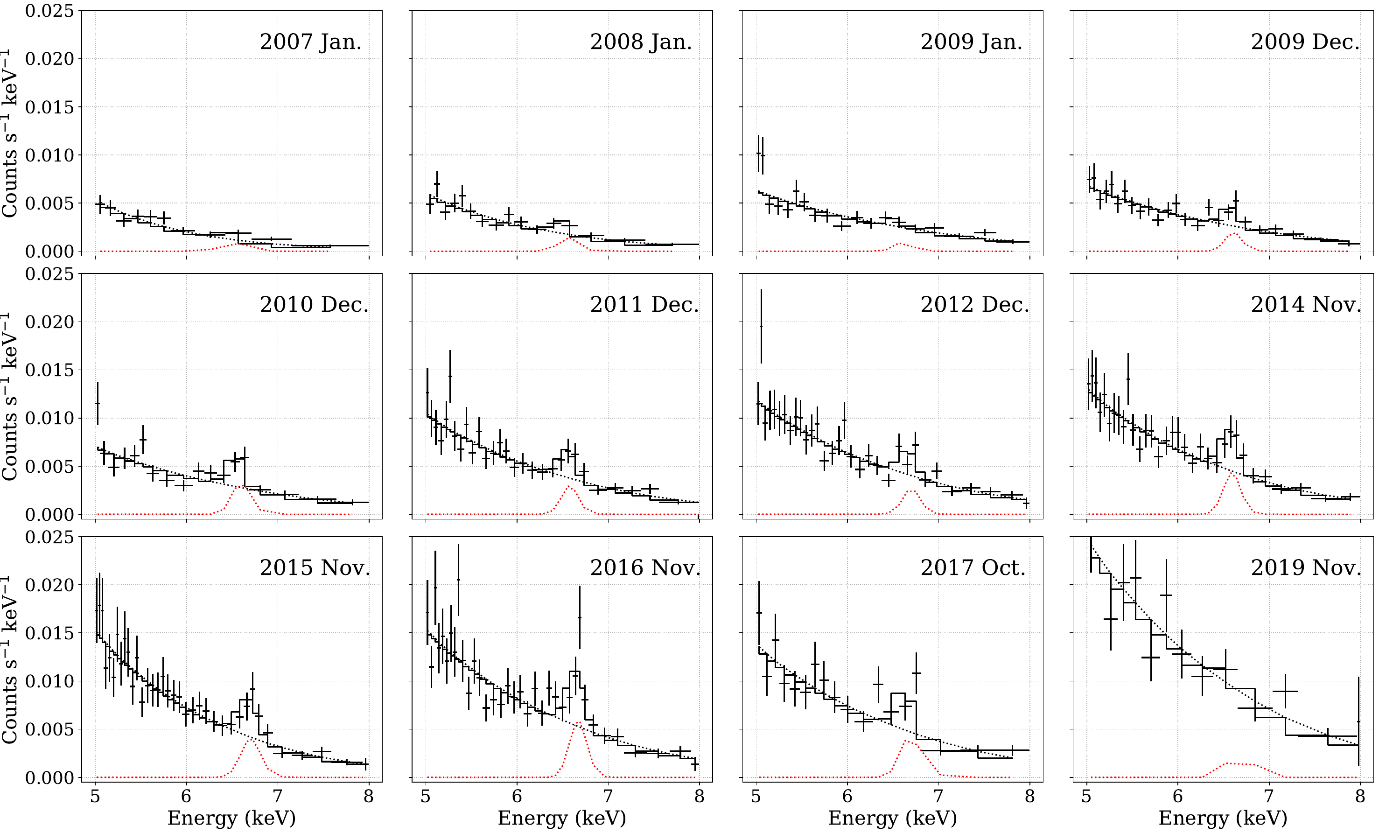}
		\caption{The $5.0$--$8.0$\,keV EPIC-pn spectra of SNR 1987A, which show the evolution of the Fe K lines. The best-fit model is plotted in black, which contains the bremsstrahlung continuum (black dashed line) and the Gaussian profile (red dashed line). \label{fig:Fe_K}}
	\end{figure*}

\begin{figure*}
		\plotone{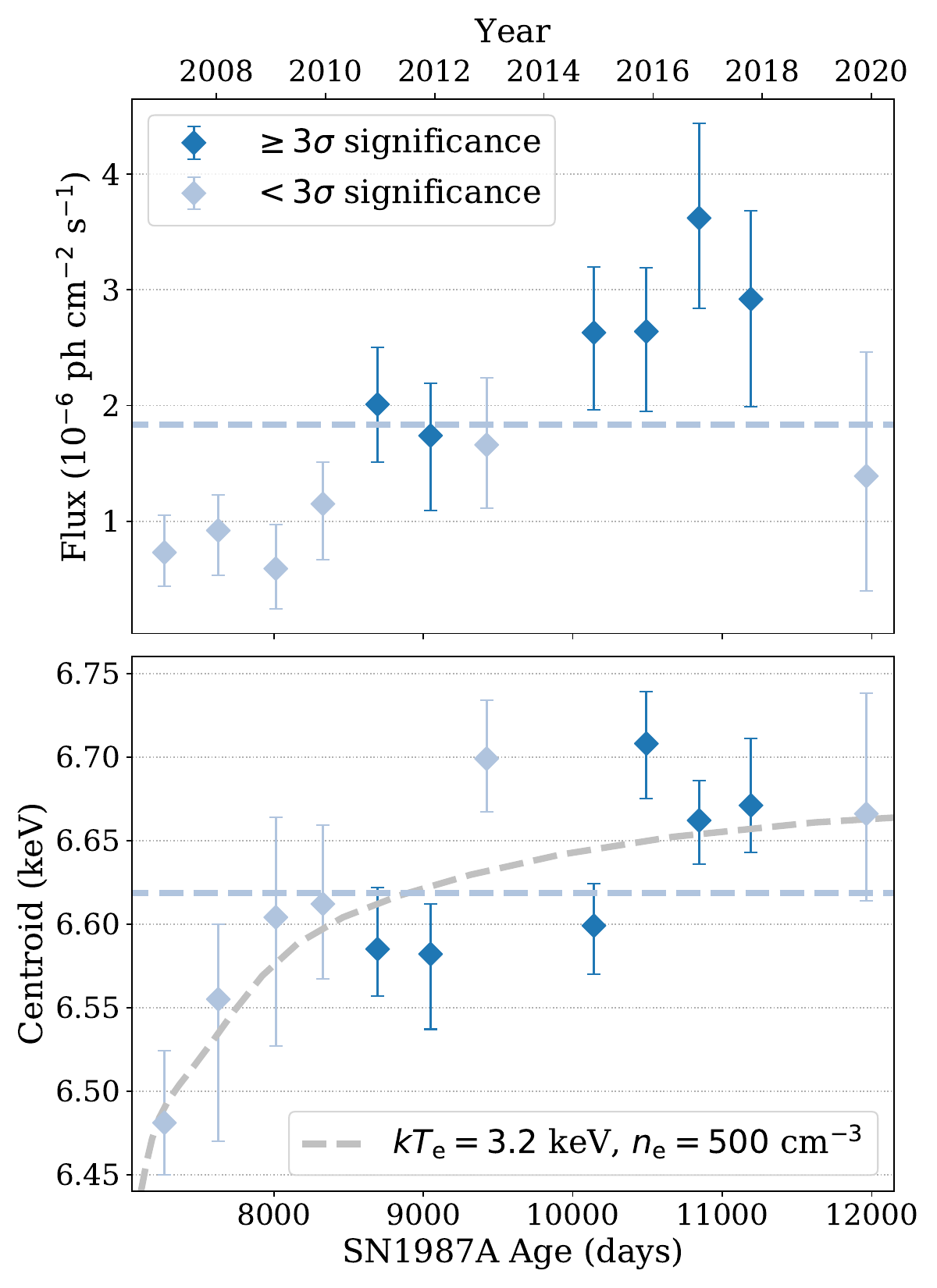}
		\caption{Top: the Fe K light curve. The points with $\gtrsim3\sigma$ significance are shown in dark blue, while those with $<3\sigma$ significance are shown in light blue. The blue dashed line denotes the mean flux. Bottom: the Fe K line centroids. The blue dashed line denotes the mean value, and the gray dashed curve shows the modeled line centroid for a plasma with $kT_{\rm e}=3.2$\,keV and $n_{\rm e}=500$\,cm$^{-3}$, shocked at 7000 days after the explosion.\label{fig:Fe_K_flux_cen}}
	\end{figure*}

	We extracted the $5.0$--$8.0$\,keV EPIC-pn spectra, and fitted them with a {combination of} a thermal bremsstrahlung continuum ({\tt bremss}) and a Gaussian profile ({\tt gauss}). As shown in Figure \ref{fig:Fe_K}, the Fe K line gradually emerged out from the underlying continuum. With the Gaussian component, we constrained the line fluxes and the centroids of the Fe K lines (listed in Table \ref{tab:Fe K_AIC}). 
	{There are several approaches to evaluate the significance of the lines. As listed in Table \ref{tab:Fe K_AIC}, we first provided the measured line fluxes and their 1-$\sigma$ uncertainties as estimates. Following \citet{2012AA...548L...3M}, we also performed F-tests and calculated the p-values. At last, we introduced the Akaike information criterion \citep[AIC,][]{1974ITAC...19..716A} as a third approach. We derived the AIC value as:
	\begin{equation}
		AIC=n\ln\left[\sum_{i}\omega_i\left(y_i-m_i\right)^2/n\right]+2p=n\ln\left(\chi^2/n\right)+2p,
	\end{equation}
	where $n$ is the number of measurements, $p$ the number of free parameters, $y_i$ the observations, $m_i$ the model predicted values, and $\omega_i$ the weight factors. The more appropriate model is the one with the smaller AIC value. Thereby, we calculated the differences in AIC values between the models with and without the Gaussian component, and listed them in Table \ref{tab:Fe K_AIC} as $\Delta AIC$.}
	{Based on the significance analysis, we found that} the first few observations taken in 2007--2009 show only modest (1--2$\sigma$) detections. But since 2010, {we have continuously detected} the Fe K lines with $\gtrsim3\sigma$ significance ---the less significant detection in 2019 
	{is likely} caused by the rather short effective exposure time of $\sim11$\,ks. Despite of a few less significant detections, we find that both the flux and the centroid energy of the Fe K lines have been increasing during the last decade, 
	{as shown in
		Figure \ref{fig:Fe_K_flux_cen}. Assuming that the flux and the
		centroid energy are constant leads to 
		reduced chi-square values $\chi_{\rm r}^2\approx4.05$ and  $\chi_{\rm r}^2\approx3.36$,  respectively. Both values of $\chi_{\rm r}^2$ are
		clearly inconsistent with a constant flux and centroid energy.}
	This {indicates} that there is an increase in the emission measure and the average ionization state of the {Fe-K emitting} plasma as function of time. 
	However, the equivalent width of the line, which is more indicative of the Fe abundance, {shows} no significant variation during the period.

	\section{Discussion}\label{sec:disc}
	\subsection{Temporal evolution of the plasma}\label{sec:temp_evo}
	Since the SN blast wave {impacted on} the main body of the ER, the X-ray emission from SNR 1987A {has} been dominated by the {shocked} gas inside the ring. The XMM-Newton observations presented in this work started after the main impact and covered a time interval over 12 years, which {helps} us to follow the post impact evolution and provide  further insights into the composition and structure of the ER.
	
\begin{figure*}
		\plotone{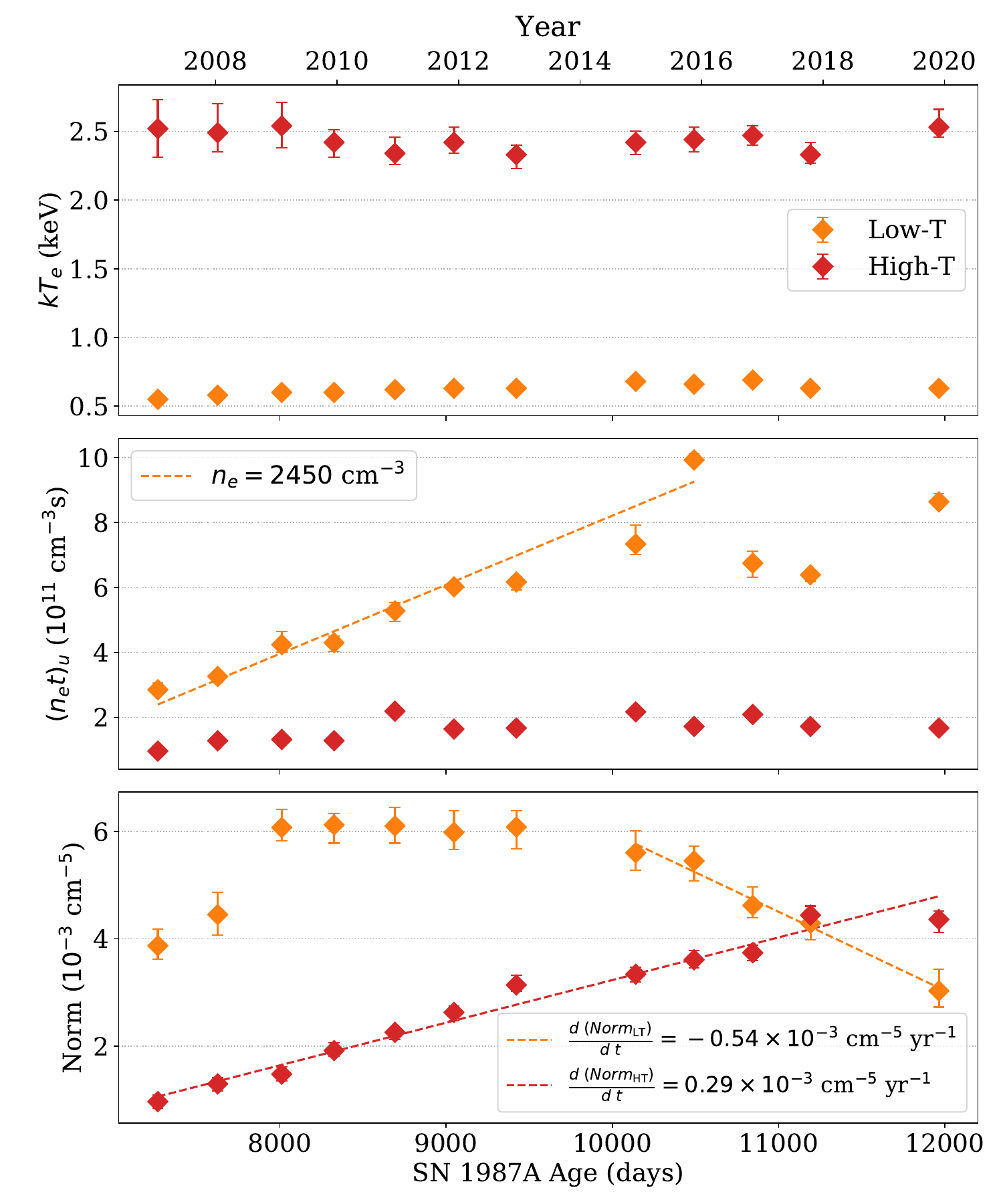}
		\caption{The electron temperatures, ionization parameters and the normalization parameters obtained in the {\tt vpshock} model fitting, with their 90\% errors. The values for the low-temperature component are denoted in orange, while those for the high-temperature components in red. In the middle panel, the orange dashed line shows a linear fit to the ionization parameters of the low-temperature component (from 2007 to 2015), corresponding to an electron density $n_{\rm e}\sim2450$\,cm$^{-3}$. In the bottom panel, the orange dashed line shows a linear fit to the normalization of the low-temperature component (from 2014 to 2019), corresponding to an decreasing rate of $\sim0.54\times10^{-3}$\,cm$^{-5}$\,yr$^{-1}$, while the red dashed line shows a linear fit to the normalization of the high-temperature component, corresponding to an increasing rate of $\sim0.29\times10^{-3}$\,cm$^{-5}$\,yr$^{-1}$.\label{fig:temp_evo}}
	\end{figure*}

	{With} the {\tt vpshock} fitting (Section \ref{sec:vpshock}), we have constrained the electron temperatures, the upper limits of the ionization parameters, and the normalizations of the gas, as shown in Figure \ref{fig:temp_evo}. 
	
	The electron temperatures of the two plasma components {show} no significant variation during the last decade ($kT_{\rm e,LT}\sim0.6$\,keV and $kT_{\rm e,HT}\sim2.5$\,keV), {indicating} basically constant shock velocities. Given the classic relation between the shock velocity and the average post shock temperature as $V_{\rm s}^2=(16/3)(\overline{kT})/(\mu m_{\rm p})$ (where $\mu$ is the average particle mass in units of the proton mass, $\mu\sim0.72$ for SNR 1987A abundance), we obtained $V_{\rm s,LT}\sim650$\,km\,s$^{-1}$ and $V_{\rm s,HT}\sim1330$\,km\,s$^{-1}$ {for the low- and high-temperature plasma, respectively}. On the other hand, \citet{2016ApJ...829...40F} constrained the expansion velocities of the ER as $\sim1850$\,km\,s$^{-1}$ for the soft band (0.5--2\,keV) and $\sim3070$\,km\,s$^{-1}$ for the hard band (2.0--10\,keV) {by fitting spatial models to Chandra ACIS images}, which also provided {estimates} of the typical shock velocities {and are generally higher than what we found here.} {This difference might further reflect the non-equilibrium of the {temperatures} between different particle species {behind the collisionless shock front}. In that case, the electron temperature could be much lower than the proton/ion temperatures or the average temperature of the plasma \citep[e.g.,][]{2020..pesnr.book}, and thus cause an underestimate of the shock velocities in the calculations above. Evidence for this temperature non-equilibrium effect have been found by both simulations and observations \citep[e.g.,][]{1988ApJ...329L..29C,2000ApJ...543L..67S,2003ApJ...587L..31V,2007ApJ...654L..69G}. Recently, \citet{2019NatAs...3..236M} measured the post-shock temperatures of protons and ions in SNR 1987A, and found that the ratio of ion temperature to proton temperature is significantly higher than one, providing observational clues to the temperature non-equilibrium in this young remnant.}
	
	We found that the upper limit on the ionization parameter of the low-temperature plasma ($\tau_{\rm u,LT}$)  has been increasing linearly from 7300 to 10500 days. {A linear fit to $\tau_{\rm u,LT}$ {suggests} an average density of $n_{\rm e,LT}=2450\pm150$\,cm$^{-3}$.} By using a similar method, \citet{2010AA...515A...5S} fitted a linear function to $\tau_{\rm u,LT}$ from 2007 to 2009 ({7300 to 8000 days}, based on the first three XMM-Newton observations listed in Table \ref{tab:obs}), {which}
	yielded a density {of} $3240\pm640$\,cm$^{-3}$. They also found that for the high-temperature plasma, $\tau_{\rm u,HT}$ has been increasing steadily from 2003 to 2009 ({6000 to 8000 days}). However, our results {indicate} that after 2009, $\tau_{\rm u,HT}$ stayed constant at around $2\times10^{11}$\,cm$^{-3}$\,s, and {show} no further evolution. Our fit to $\tau_{\rm u,LT}$ also {indicates} that most of the low-temperature gas {was} first shocked at around 6000 days after the explosion, which {gives} us an estimate of the date of impact {on} the main ER.  This is {consistent} with {the} previous results based on X-ray light curve analysis and expansion velocity studies \citep[e.g.,][]{2009ApJ...703.1752R,2013ApJ...764...11H,2016ApJ...829...40F}.
	
	The normalization parameters of the two plasma components {show} different variations during the last decade. For the low-temperature plasma, {the normalization} ${\rm Norm_{LT}}$ increased at the first few years, leveled off at around 9000 days, and then started to decrease steadily since $\sim10500$ days. On the other hand, ${\rm Norm_{HT}}$ kept increasing during the whole interval, with an approximately constant rate. {Linear fits to the data gave a decreasing rate of ${\rm Norm_{LT}}$ (after 10500 days) as $(0.54\pm0.03)\times10^{-3}$\,cm$^{-5}$\,yr$^{-1}$, and an increasing rate of ${\rm Norm_{HT}}$ as $(0.29\pm0.01)\times10^{-3}$\,cm$^{-5}$\,yr$^{-1}$.}
	
\begin{figure*}
		\plotone{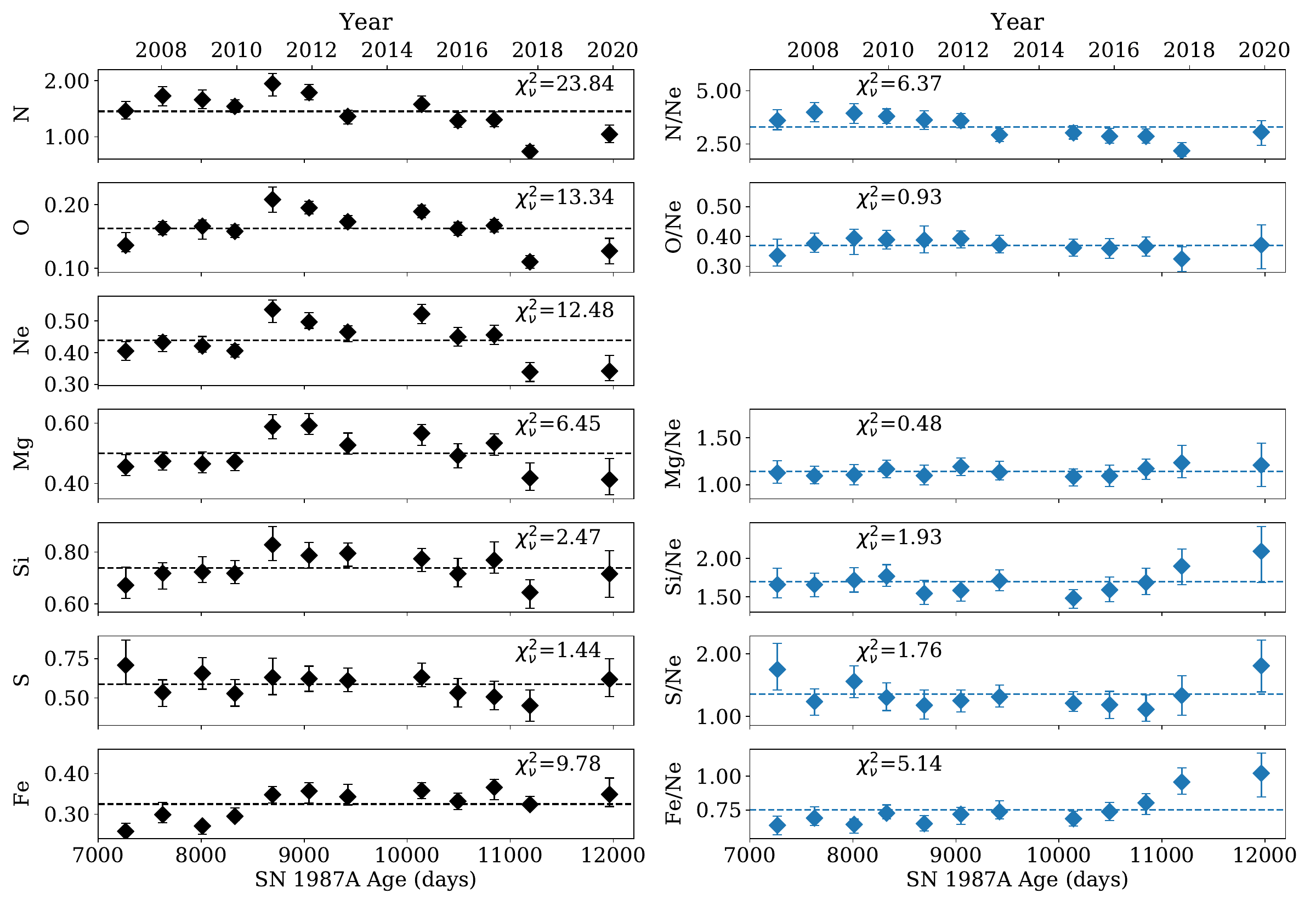}
		\caption{The abundances of different metal species (left), and the abundance ratios with respect to Ne (right). The black/blue dashed lines indicate the mean values. Error bars represent the 90\% uncertainties.\label{fig:abund}}
	\end{figure*}

	With the spectral analysis, we have also constrained the abundances of {various} metal species, including N, O, Ne, Mg, Si, S, and Fe, as shown in {the} left panels of Figure \ref{fig:abund}. We note that the metal abundances, {which} are tied between two plasma components in our spectral fitting, should be taken as the (weighted) average values. We also calculated the abundance ratios with respect to Ne (Ne abundances are best constrained among all species), {as shown in the} right panels of Figure \ref{fig:abund}. In order to investigate their variations during the period, we derived the mean abundances and abundance ratios, and performed $\chi^2$ analysis based on these mean values (shown in Figure \ref{fig:abund}). We note that for a p-value of 0.002 ($\sim3\sigma$ level), the critical reduced chi-square {is given as} $\chi^{2}_{\rm \nu,c}\approx2.67$ for 11 degrees of freedom. {Thus the $\chi^2$ analysis {indicates} that the {abundances} of N, O, Ne, Mg, and Fe, as well as the N/Ne and Fe/Ne ratio, have significantly changed in the last decade.} The variations of N, O, Ne, and Mg abundances follow a similar track, which stayed at their mean values till around 10500 days, and then started to decrease recently. This is well fitted with the recent decline of the low-temperature plasma normalization after $\sim10500$ days. {The increasing Fe abundance conforms to the increasing high-temperature plasma normalization. As for the abundance ratios, N/Ne has been decreasing while Fe/Ne been increasing during the time, consistent with the variation of single abundances.} {Therefore, the overall changes of the single metal abundances and their ratios could be naturally explained {with} the different chemical compositions {for} 
	two plasma components.} In our spectral fitting, the abundances of N, O, Ne, and Mg are determined mainly by their emission lines lying in 0.5--2\,keV energy band, which were dominated by the low-temperature component at the beginning. Since the recent decline of ${\rm Norm_{LT}}$, the high-temperature plasma started to contribute a significant part of the emission even in the soft band, and thus the abundances became more representative of the high-temperature plasma. {In that case, our results {indicate} that the low-temperature plasma has enhanced metal abundances {as compared to} the high-temperature plasma, especially for N, O, Ne, and Mg, while the high-temperature plasma may be more abundant in Fe. We will further discuss the possible origin of metal differences in Section \ref{sec:origin}. It is also possible that the increasing Fe abundance and the Fe/Ne ratio result from the recent emergence of Fe-rich ejecta. However, this needs to be further examined, and we leave the discussions in Section \ref{sec:Fe_K_origin}.} {At last, }we note that in a recent paper, \citet{2020ApJ...899...21B} studied the metal abundances of SNR 1987A based on Chandra HETG data (taken in 2011--2018) and found no significant variations during the period. However, the spectra they adopted were limited in 0.5--3 keV, thus the hot component might not be well constrained. Specifically, they obtained a temperature of 1.2-1.6 keV for the hot plasma, while our results and many other works suggested $\sim2.5$ keV. 
	
	\subsection{Electron density, ionization age, and filling factor}\label{sec:ave_pro}
	In this section, we estimate the electron density, the ionization age, and the filling factor of the X-ray emitting gas based on spectral fitting results.
	
	Assuming that the {shocked} ER has a toroidal geometry, its volume can be derived as $V_{\rm ER}=2\pi^2Rr^2d^3$, where $R$ and $r$ are the angular radius and the half-width of the torus, {respectively}, and $d$ is the distance to SNR 1987A. We then took the total X-ray emitting volume as $V=f_{\rm tot}V_{\rm ER}$, where $f_{\rm tot}$ is the total filling factor. With this volume given, the electron density can be {derived from} the normalization parameter ${\rm Norm}=10^{-14}/(4\pi d^2)n_{\rm e}n_{\rm H}V$, where $n_{\rm H}\approx n_{\rm e}/1.2$ for a fully ionized plasma with near solar abundance. The ionization age $t$ can then be {obtained} from $\tau=n_{\rm e}t$. By further assuming that the two plasma components are in pressure balance (i.e., $n_{\rm e,LT}kT_{\rm e,LT}=n_{\rm e,HT}kT_{\rm e,HT}$) and share the whole X-ray emitting volume (i.e. $f_{\rm tot}=f_{\rm LT}+f_{\rm HT}$), we were able to constrain these parameters {separately} for each component.
	
	The geometry parameters $R$ and $r$ of the {shocked} ER in different epochs have been constrained by fitting a ``torus $+$ lobes'' spatial model to the deconvolved Chandra images \citep[e.g.,][]{2009ApJ...703.1752R,2016ApJ...829...40F}. Here, we adopt a half-width $r=0.22''$ for all the observations according to \citet{2009ApJ...703.1752R}, who found a nearly constant ring width after $\sim6000$ days. On the other hand, the average radius $R$ is increasing due to the expansion of the blast wave. For each observation, we set $R$ to be the value obtained by \citet{2016ApJ...829...40F} using the Chandra data taken in a similar epoch ({for the observations taken after 2015, $R$ were estimated by an extrapolation based on the expansion velocity}).  We adopt a distance to SNR 1987A of $d=51.4$\,kpc based on \citet{2005coex.conf..585P}. We use the {\tt vpshock} fitting results for calculations. As for ionization ages, we {estimated} both the average $t$ {from} the average $\tau$ obtained in {the} {\tt vnei} fit, and the upper limit $t_{\rm u}$ {from} $\tau_{\rm u}$ obtained in {the} {\tt vpshock} fit. The derived electron densities, average/maximum ionization ages, and filling factors are listed in Table \ref{tab:derived_par_nH_fixed}.
	
	{The electron densities of the two plasma components {are derived} 
	to be $n_{\rm e,LT}\sim(1600$--$2700)f^{-0.5}_{\rm tot}$\,cm$^{-3}$ (with an average value $\sim2180f^{-0.5}_{\rm tot}$\,cm$^{-3}$) and $n_{\rm e,HT}\sim(360$--$700)f^{-0.5}_{\rm tot}$\,cm$^{-3}$ (with an average value $\sim550f^{-0.5}_{\rm tot}$\,cm$^{-3}$).} The unknown filling factor $f_{\rm tot}$ led to some uncertainties in our results. Ideally, it should be around 1 if the total X-ray emitting volume is comparable to the {shocked} ER volume. However, it will be significantly lower than 1 if the X-ray emitting gas is  clumpy and concentrates in a small part of the {shocked} ring. It could also go beyond 1 if the hot, lower-density, gas extends to a larger region outside the ring. We find that the derived gas densities appear to be increasing during the period, which may be indicative of a density gradient {throughout} the ring. But this is still {debatable due to uncertainties in the time-variable filling factor and the assumed geometry.} 
	{As mentioned above in Section \ref{sec:temp_evo}, we got a density of $n_{\rm e,LT}=2450\pm150$\,cm$^{-3}$ for the low-temperature plasma by fitting a linear function to $\tau_{\rm u,LT}$.} {As a comparison}, \citet{2010AA...515A...5S} obtained the average densities as $n_{\rm e,LT}=3240\pm640$\,cm$^{-3}$ and $n_{\rm e,HT}=480\pm140$\,cm$^{-3}$ for the two components {from} a similar analysis of the 2003--2009 XMM-Newton observations. On the other hand, expansion-velocity studies suggest a shock velocity of $\sim6800$\,km\,s$^{-1}$ before {the} impact, while it dropped to $\sim1850$\,km\,s$^{-1}$ {as viewed} in 0.5--2\,keV band and $\sim3070$\,km\,s$^{-1}$ in 2--10\,keV band after 6000 days \citep[e.g.,][]{2016ApJ...829...40F}. This {requires} that the density increased by a factor of $\sim13$ and $\sim5$ {(given the relation between the shock velocity and the gas density as $v_{\rm s}\propto n_{\rm e}^{0.5}$)}. Given that the \ion{H}{2} region has a density $\sim10^2$\,cm$^{-3}$, this {indicates} an average density $\gtrsim10^3$\,cm$^{-3}$ for the soft band emitting gas and $\sim500$\,cm$^{-3}$ for the hard band emitting gas. {Taking} all these results together, we {suggest} $n_{\rm e,LT}\sim2400$\,cm$^{-3}$ and $n_{\rm e,HT}\sim550$\,cm$^{-3}$ as reasonable estimations of the low- and high-temperature plasma densities.
	
	The ionization age characterizes the elapsed time since the gas was shocked. The average ionization age $t$ and its upper limit $t_{\rm u}$ were constrained based on {the} {\tt vnei} and {\tt vpshock} spectral fitting results, {respectively}. For the low-temperature plasma, we found $t_{\rm u,LT}$ gradually increased from $\sim2000f^{0.5}_{\rm tot}$\,days to $\sim4800f^{0.5}_{\rm tot}$\,days during the period of 7300--10500 days after explosion, {consistent} with the physical picture that the ER started to be shocked by the blast wave at $\sim6000$ days and dominated the soft band X-ray emission since then. Under the plane-parallel shock assumption, we would expect the average ionization age $t_{\rm LT}=t_{\rm u,LT}/2\sim(1000$--$2400)f^{0.5}_{\rm tot}$\,days. However, it is found to be significantly lower, {as indicated by the {\tt vnei} fitting}, with $t_{\rm LT}\sim(300$--$500)f^{0.5}_{\rm tot}$\,days, and {does not show evidence for a similar increase as} 
	$t_{\rm u,LT}$. 
	This {may suggest} that the low-temperature component has always been dominated by the ``youngest'' hot gas, which has just been shock heated within the most recent 1--2 yr. This very dynamic shock-heating process {is} also indicated by Chandra imaging observations. As shown by \citet[][Figure 5 {therein}]{2016ApJ...829...40F}, the surface X-ray brightness distribution of the ER can significantly change within a few hundred days, which means the brightening and the fading of the gas take place on a rather short timescale. This short average ionization age may be related to the shock transmission timescale and/or the destruction timescale of the shocked {high-density clumps.}
	{However, it should be remembered that the {\tt vnei} model fitting optimizes for the line emission, and the line emissivity has a non-linear relation with the ionization state. As a result, the line emission is suppressed when the gas is close to equilibrium at these temperatures, so the {\tt vnei} parameters can be skewed to lower $\tau$ values.}
	We have also considered whether the short timescale {could reflect the radiative cooling timescale}, which can be estimated as $\tau_{\rm cool}\approx5.7kT/(n_{\rm e}\Lambda(T))$, {where $\Lambda$ is the volumetric cooling rate} \citep[e.g.,][]{2020..pesnr.book}. 
	{For} the low-temperature plasma, {taking} $kT_{\rm e,LT}\sim0.6$\,keV, $n_{\rm e,LT}\sim2400$\,cm$^{-3}$ and $\Lambda(T)\sim10^{-22}$\,erg\,s$^{-1}$\,cm$^{-3}$ \citep[e.g.,][]{2009AA...508..751S}, we obtained a cooling timescale of $\tau_{\rm cool,LT}\sim720$\,yr, which {makes} radiation cooling unlikely to be the cause of the short average ionization age. 
	{The optical emission is likely to come from even denser clumps ($\gg 10^4$~cm$^{-3}$), for which the post-shock temperature is in a regime for which $\tau_{\rm cool}$ is an order of magnitude smaller, with time scales of months to years. }
	
	{For the high-temperature plasma, we obtained $t_{\rm u,HT}\sim(2800$--$4600)f^{0.5}_{\rm tot}$\,days (with an average value $\sim3400f^{0.5}_{\rm tot}$\,days), which {shows} no significant variation during this period. The average ionization age {is derived to be}  $t_{\rm HT}\sim(1000$--$1500)f^{0.5}_{\rm tot}$\,days ---with an average value $\sim1300f^{0.5}_{\rm tot}$\,days--- 
	{consistent} with the expected value under the plane-parallel shock assumption.}
	
	The filling factors have also been constrained 
	for the two plasma components, respectively. The ratio $f_{\rm LT}/f_{\rm HT}$, which is no longer depending on the uncertain total filling factor $f_{\rm tot}$, 
	{is} found to have been $\sim0.2$ during 7300--9000 days, and it has been continuously decreasing since then. By the latest observation at $\sim12000$ days, it has dropped to $\sim0.04$.
	
	{With} the derived electron densities and filling factors, we also estimated the gas masses for each of the components, as listed in Table \ref{tab:derived_par_nH_fixed}.
	
	\subsection{Origin of the X-ray emitting plasma}{\label{sec:origin}}
	Similar to many previous studies, our work {reveals} that the X-ray spectra of SNR 1987A can be approximately described by the thermal emission from a combination of two major plasma components, one with a low temperature $\sim0.6$\,keV, and the other one with a high temperature $\sim2.5$\,keV. However, the physical origins of these plasmas could be rather complicated due to the complex CSM structure and {the} shock system. The main structure of the shocked CSM includes the ER, an ionized \ion{H}{2} region surrounding the ring, and the high-latitude material beyond the ring. The ER consists of several dense clumps (``fingers'') distributed mainly in its inner portion and the smooth inter-clump gas. It has been suggested that the low-temperature plasma (or the soft X-ray emission) comes from the slow transmitted shock into the dense clumps, while the high-temperature plasma (or the hard X-ray emission) mainly comes from the less-decelerated shocks (which may also {include  reflected shocks}) moving through the lower density {ring} material. This physical picture was first indicated by the observations taken in the early stage of the shock-ring interaction \citep[e.g.,][]{2006ApJ...646.1001P,2010MNRAS.407.1157Z}, and later supported by {the} 1D to 3D hydrodynamic simulations \citep[e.g.,][]{2012ApJ...752..103D,2015ApJ...810..168O}. However, our analysis {is} based on the more recent observations (after $\sim7500$ days), thus {represents} the later stage evolution. Keeping this in mind, we {revisit here} the origin of the X-ray emitting plasma.
	
	Based on the early optical analysis of the ionized gas in the ER, the dense clumps have a density $\sim$(1--3)$\times10^{4}$\,cm$^{-3}$, and the {inter-clump gas} has a density $\sim$(1--6)$\times10^{3}$\,cm$^{-3}$ \citep[e.g.,][]{1996ApJ...464..924L,2010ApJ...717.1140M}. Similar densities {are} also indicated by hydrodynamic simulations \citep[e.g.,][]{2012ApJ...752..103D,2015ApJ...810..168O}. We {find that} 
	the low-temperature plasma has a density of $n_{\rm e,LT}\sim2400$\,cm$^{-3}$, {and} thus {it is}
	more indicative of the lower density gas inside the ring. It may consist of both the shocked {inter-clump gas} and the evaporated clump material (or the fragments with lower density). On the other hand, the high-temperature plasma has a density of $n_{\rm e,HT}\sim550$\,cm$^{-3}$, which is lower than the expected density of the ring material. However, {taking into account} 
	a shock-compression ratio of $\sim4$, it is consistent with the interpreted density of the \ion{H}{2} region \citep[$\sim10^{2}$\,cm$^{-3}$, e.g.,][]{1996ApJ...464..924L,2010ApJ...717.1140M,2015ApJ...810..168O}. 
	
	The shocked dense clumps, which have a typical density of $\gtrsim10^4$\,cm$^{-3}$, are expected to contribute a rather low temperature ($\lesssim0.1$\,keV, given the density ratio between the dense clump and the {inter-clump} material) plasma component. However, we have not found solid evidence for such a component in our spectral analysis. We note that at the early stage of the shock-ring interaction, the soft (low-temperature) component was found to be close to collisional ionization equilibrium with a much lower electron temperature $\sim0.22$\,keV, the density of which was estimated as $\sim7500$\,cm$^{-3}$ \citep{2004ApJ...610..275P}. As the blast wave penetrated further into the ER, the temperature of this soft component gradually increased to $\gtrsim0.5$\,keV \citep[e.g.,][]{2006ApJ...646.1001P,2010AA...515A...5S}. This {indicates} that the soft emission was dominated by the shocked dense clumps at the beginning, but was later overtaken by the lower density gas inside the ring. The dense clumps themselves could either have faded out due to evaporation/destruction, or {they contribute} little to the overall emission, which can hardly be detected.
	
	The total mass of the ER was {estimated} to be $\sim0.06$\,$\Msun$ {from} optical observations and hydrodynamic simulations \citep[e.g.,][]{2010ApJ...717.1140M,2015ApJ...810..168O}, and most of the mass {is distributed} in the {lower density} component ($\sim$60\%--80\%). Despite the large uncertainty of our mass estimate, it provides hints to the plasma origin. {We found a peak mass value for the low-temperature plasma $\sim0.08f_{\rm tot}^{0.5}$\,$\Msun$}, which is {similar to} the total mass of the ER. On the other hand, the mass of the high-temperature plasma has been continuously increasing, and was found to be $\gtrsim0.15f_{\rm tot}^{0.5}$\,$\Msun$ {in} latest observations, which {exceeds} the expected ER mass by a factor of 2. Thereby, the low-temperature plasma could be dominated by the shocked ring material, while the high-temperature plasma should also contain the material around/beyond the ring, which may extend to a larger radius and has been continuously shocked by the blast wave.
	
	As described in Section \ref{sec:temp_evo}, the metal abundances of N, O, Ne, and Mg have significantly declined in the recent few years, which may result from the different chemical compositions between two plasma components. In that case, it is more reasonable to interpret the two plasma components as representing the different parts of the CSM structure. The ER was found to have enhanced He and N abundances according to the optical observations \citep[e.g.,][]{1996ApJ...464..924L,2010ApJ...717.1140M}. {As the enhancement in He and N implies a processing of hydrogen, it affects the abundances of alpha- and Fe-group elements with respects to hydrogen.}
	The \ion{H}{2} region, as indicated by the early X-ray observations taken before the impact, has relatively lower abundances of N, O, Ne, and Mg \citep{2002ApJ...567..314P}. {In} the hydrodynamics-based analysis of the X-ray spectra, \citet{2012ApJ...752..103D} found different metal abundances in the ER and the \ion{H}{2} region, where the ER had generally higher abundances for most of the species. Our results suggested that the low-temperature plasma dominated the soft band emission until $\sim10500$ days, and resulted in enhanced average abundances of N, O, Ne, and Mg during this period. As the decline of the low-temperature emission measure, the average abundances of N, O, Ne, and Mg started to decrease in the last few years, which in turn indicated the lower metal abundances of the high-temperature plasma. Therefore, the low-temperature plasma could be more representative of the ER, while the high-temperature plasma could be more representative of the \ion{H}{2} region and the high-latitude material with low metal abundances.
	
	Given all the discussions above, our results {indicate} that the low-temperature plasma is dominated by the lower density gas inside the ER, which may include the smooth {inter-clump gas} and the evaporated/destructed clump material, while the high-temperature plasma is dominated by the \ion{H}{2} region and the low density gas around/beyond the ring. The recent decrease in the low-temperature emission measure {indicates}
	that the blast wave has left the main ER, while it is still propagating into the high-latitude material, resulting in the continuously increasing high-temperature emission measure. Such a physical picture is also reflected by the drop of the filling factor ratio $f_{\rm LT}/f_{\rm HT}$. 
	However, the real CSM structure must be rather complicated, and the actual density and temperature distributions cannot be fully represented by a simple two-component model (below we discuss the need for at least one additional component to fit the Fe K line emission). 
	Alternative approaches could be the differential emission measure analysis based on the high resolution spectra \citep[e.g.,][]{2006ApJ...645..293Z,2009ApJ...692.1190Z}, and detailed full-scale HD/MHD simulations \citep[e.g.,][]{2015ApJ...810..168O,2019AA...622A..73O,2020AA...636A..22O}.

	\subsection{Possible origins of the Fe K lines}\label{sec:Fe_K_origin}
\begin{figure*}
		\plotone{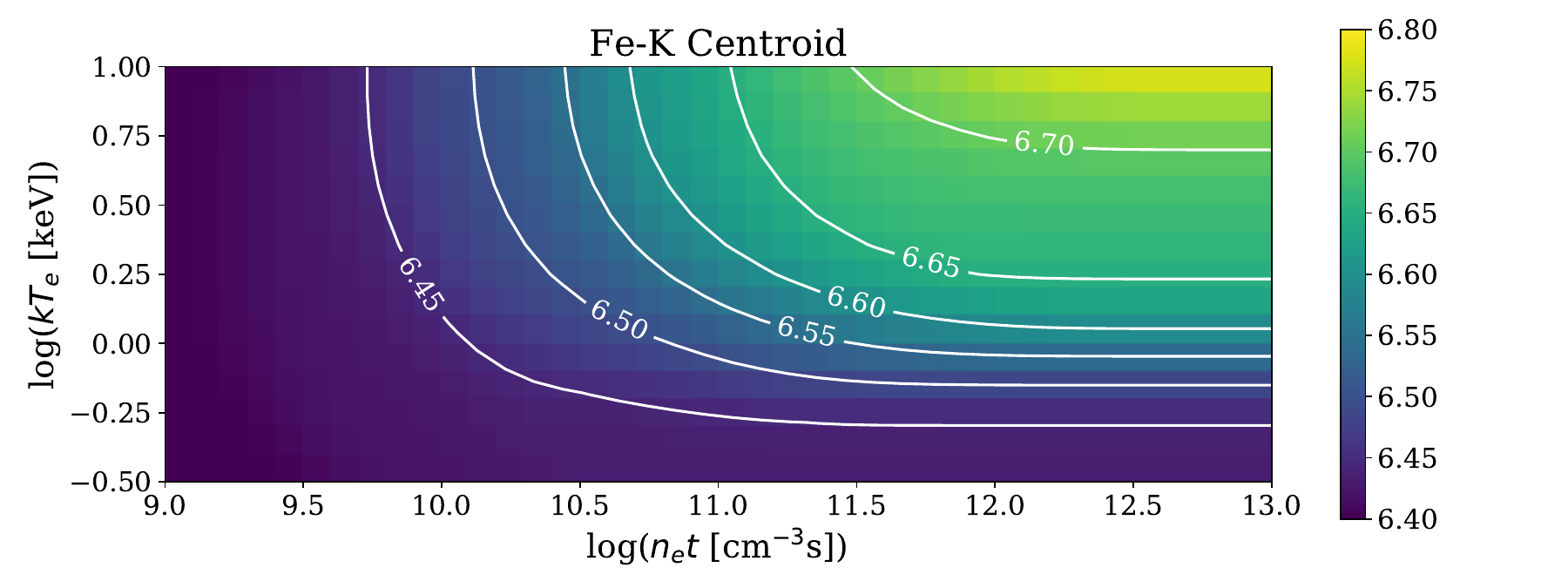}
		\caption{Theoretical predicted Fe K centroid energies, as a function of ionization parameter and electron temperature.\label{fig:Fe_K_diagram}}
	\end{figure*}

	{The EPIC-pn camera  reveals} Fe K features in SNR 1987A's spectra, the origin of which remains unclear. Recent observations {indicate} a Fe K centroid energy $\gtrsim6.65$\,keV. By comparing with the {theoretically} predicted line centroid shown in Figure \ref{fig:Fe_K_diagram} \citep[based on SPEX code\footnote{https://www.sron.nl/astrophysics-spex},][]{1996uxsa.conf..411K}, the high centroid energy {requires} either a high plasma temperature $\gtrsim2.0$\,keV or a large ionization parameter $\gtrsim10^{11}$\,cm$^{-3}$\,s. 
	{Since the Fe K line emission is not well reproduced with the standard two-temperature component scenario,}
	we {invoked}
	a third, hot plasma component as the origin of the Fe K lines. Taking the 2015 observation as an example, we refit the RGS and pn spectra with a three-temperature {\tt vnei} model. {It provides 
	an overall 
	{improved} fit ($\chi_{\rm r}^2\sim1.19$, versus $\chi_{\rm r}^2\sim1.33$ for a two component fit, or $\Delta \chi^2=320$ for three additional degrees of freedom) to the data and reproduces Fe K lines successfully. The Fe K features are 
	characterized by a high-temperature plasma component with $kT_{\rm e}=2.9^{+0.2}_{-0.1}$\,keV and $\tau=(1.4\pm0.1)\times10^{11}$\,cm$^{-3}$\,s. The other two components have lower temperatures and similar ionization parameters: $kT_{\rm e}=0.35^{+0.04}_{-0.01}$\,keV, $\tau=(3.2\pm0.1)\times10^{11}$\,cm$^{-3}$\,s for the low-temperature one, and $kT_{\rm e}=0.87^{+0.03}_{-0.02}$\,keV, $\tau=1.9^{+0.3}_{-0.2}\times10^{11}$\,cm$^{-3}$\,s for the mid-temperature one.} On the other hand, the temporal variation of the Fe K centroid energy provided further information {about} the plasma properties. Given a certain temperature and density of the plasma, the relation shown in Figure \ref{fig:Fe_K_diagram} results in a modeled curve of Fe K centroid as a function of time, which can be compared with observations. {As shown in} Figure \ref{fig:Fe_K_flux_cen}, the observed Fe K centroid energies basically match with the predicted evolution of a plasma with $kT_{\rm e}\sim 3.2$\,keV and $n_{\rm e}\sim 500$\,cm$^{-3}$, shocked at 7000 days after the explosion. This Fe-K emitting component may thereby also come from the low density gas around/beyond the ER, {albeit} 
	with an even higher temperature. It may be related to the reflected shock which further compresses and heats the already shocked gas \citep[e.g.,][]{2009ApJ...692.1190Z,2010MNRAS.407.1157Z}. It is also possible that the Fe K emission comes from one or several  reverse shocked Fe-rich ejecta clumps, as the reverse shock with high ejecta-frame velocity could result in high post-shock temperature. Although most of the Fe-rich ejecta may concentrate in the inner part of the ejecta material, there could still be some high-speed debris which has already been approached by the reverse shock. 
	{We note that the gamma-ray data taken a few months after the {SN} explosion  provided surprising evidence for fast core-ejecta material mixing into the outer envelopes \citep[see][and references therein]{1990ApJ...357..638L}.}

	\section{Conclusion}\label{sec:con}
	Since entering the remnant phase, SNR 1987A has been dynamically evolving on timescales of months to years. In this work, we utilized the XMM-Newton observations of SNR 1987A from 2007 to 2019 to investigate its post-impact evolution. The high energy resolution of RGS helped us identify and measure individual emission lines, as well as the EPIC-pn camera provided us {with} an overall energy coverage from 0.3 to 10 keV and thus a better constraint of the hard band emission. We performed detailed modeling to the spectra, constrained the electron temperatures, the ionization parameters, and the emission measures of the plasma, and tracked their recent evolution. We summarize the main results and conclusions below.
	
	\begin{enumerate}
		\item The post-impact X-ray emission of SNR 1987A can be approximately described {with} a model containing two NEI plasma {components} with different electron temperatures ($kT_{\rm e,LT}\sim0.6$\,keV and $kT_{\rm e,HT}\sim2.5$\,keV). The plasma temperatures {showed} no significant variations during the last decade.
		\item The {emission measure} of the low-temperature plasma started to decrease since around 10500 days after the explosion, resulting in the recent decline of the soft band (0.5--2.0\,keV) flux. On the other hand, the high-temperature emission measure, as well as the hard band (3.0--8.0\,keV) flux, kept increasing with a steady rate.
		\item The average abundances of N, O, Ne, and Mg have decreased in the last few years, which could {result from} the different chemical compositions between two plasma components.
		\item The densities of the X-ray emitting {gases} were estimated as $n_{\rm e,LT}\sim2400$\,cm$^{-3}$ and $n_{\rm e,HT}\sim550$\,cm$^{-3}$ for the low-temperature and the high-temperature components, respectively.
		\item The low-temperature plasma is indicated to be dominated by the lower density gas inside the ER, which may include the {inter-clump gas} and the evaporated/destructed clump material. The high-temperature plasma could be dominated by the \ion{H}{2} region and the low density gas around/beyond the ring. The blast wave has now left the main ER, but is still propagating into the high-latitude material.
		\item 
		{Fe K line emission has} been detected in most of the observations. {Its} centroid energy has been increasing during the last decade, with a latest value of $\gtrsim6.65$\,keV. The Fe K lines may originate from a third, high-temperature component, which could be the reflected shock heated CSM, or be related to the reverse shocked Fe-rich ejecta.
	\end{enumerate}
	
	SNR 1987A has now entered a new evolutionary stage, {in which} the blast wave {is} starting to shock the high-latitude material beyond the ER. On the other hand, as {the} reverse shock penetrating deeper into the inner ejecta, 
	{the shocked ejecta}
	will {start to} contribute to the X-ray emission as well. Further observations of the remnant in the next few years will be crucial to {monitor} these new changes and unveil the underlying physics.
	
	\section*{note added in proof}
	{During the refereeing process of the present paper, two new analyses regarding the X-ray emission from SN 1987A were presented \citep{2021arXiv210109029G,2021arXiv210302612A}, incorporating NuSTAR
	hard X-ray spectra. 
	Both papers agree that there is a need for a third hard X-ray component, but the \cite{2021arXiv210302612A} paper (based on XMM-Newton RGS + NuSTAR data) attributes this third component to thermal emission, and the \citet{2021arXiv210109029G}
	paper (based on Chandra + NuSTAR data) attributes it to non-thermal emission from a heavily absorbed emerging pulsar wind nebula. The results presented by us are indicative of a third, high-temperature thermal component, and thus qualitatively agree with \cite{2021arXiv210302612A}. However, they found a temperature as high as $\sim4$\,keV is needed in order to fit the NuSTAR data, which is significantly higher than what we got in Section \ref{sec:Fe_K_origin}. In fact, the temperature and ionization parameter we got for the third component are closer to those obtained by \citet{2021arXiv210109029G} \citep[see also the analysis by][]{2021ApJ...907..117T} for their high-temperature thermal component. We note that our decision to include a third thermal component was based exclusively on the soft X-ray data (below 10\,keV), specifically the detection of Fe K emission, which could not be reproduced by the conventional two-component model. On the other hand, our analysis, based solely on XMM-Newton observations, cannot rule-out, or confirm the presence of a highly absorbed non-thermal component. We verified that adding an absorbed power-law component as described in \citet{2021arXiv210109029G} does not affect much the soft X-ray spectrum (producing no significant X-ray emission below 7\,keV), and thus makes no significant improvement to the fit. It remains then to be investigated, whether a fourth component representing the non-thermal emission, is still required if we would augment our analysis by incorporating hard X-ray data. But it is beyond the scope of this paper to make strong conclusions regarding this point.}
	
	\acknowledgments
	This work is partly supported by National Key R\&D Program of China No.\ 2017YFA0402600 and the NSFC under grants 11773014, 11633007, and 11851305. This project has also received funding from the European Union's Horizon 2020 research and innovation programme under grant agreement No 101004131. L. S.\ acknowledges the financial support of the China Scholar Council (No.\ 201906190108).
	\software
	{XSPEC} \citep{1996ASPC..101...17A}, SPEX \citep{1996uxsa.conf..411K}, SAS \citep{2004ASPC..314..759G}, DS9\footnote{http://ds9.si.edu/site/Home.html} \citep{2003ASPC..295..489J}

	\clearpage
	
	\begin{deluxetable*}{lll|cc|cc}
		\tablecaption{Observations\label{tab:obs}}
		\tablenum{1}
		\tablehead{
			\multicolumn{3}{c}{}&\multicolumn{2}{c}{$t_{\rm exp}$ (ks)\tablenotemark{a}}&\multicolumn{2}{c}{$\sum$GTI (ks)\tablenotemark{b}}\\
			ObsID&Date&Age (days)&EPIC-pn&RGS&EPIC-pn&RGS
		}
		\startdata
		0406840301 & 2007 Jan 17 & 7267 & 106.9 & 111.3 & 61.1 & 109.8 \\
		0506220101 & 2008 Jan 11 & 7627 & 110.1 & 114.3 & 70.7 & 102.4 \\
		0556350101 & 2009 Jan 30 & 8012 & 100.0 & 101.9 & 66.4 & 101.8 \\
		0601200101 & 2009 Dec 11 & 8327 & 89.9 & 91.8 & 82.4 & 91.7 \\
		0650420101 & 2010 Dec 12 & 8693 & 64.0 & 65.9 & 52.7 & 65.9 \\
		0671080101 & 2011 Dec 2 & 9048 & 80.6 & 82.5 & 64.2 & 80.6 \\
		0690510101 & 2012 Dec 11 & 9423 & 68.0 & 69.9 & 59.4 & 69.8 \\
		0743790101 & 2014 Nov 29 & 10141 & 78.0 & 79.6 & 56.4 & 79.4 \\
		0763620101 & 2015 Nov 15 & 10492 & 64.0 & 65.9 & 58.0 & 65.8 \\
		0783250201 & 2016 Nov 2 & 10845 & 72.4 & 74.3 & 50.3 & 74.2 \\
		0804980201 & 2017 Oct 15 & 11192 & 77.5 & 79.4 & 27.8 & 79.3 \\
		0831810101 & 2019 Nov. 27 & 11964 & 32.4 & 34.9 & 11.1 & 34.8 \\
		\enddata
		\tablenotetext{a}{Total exposure times.}
		\tablenotetext{b}{Total good time intervals after background flare removal.}
	\end{deluxetable*}

	\begin{longrotatetable}
		\begin{deluxetable*}{llllllllllllll}
			\tablecaption{Identified lines in RGS spectra\label{tab:line}}
			\tablenum{2}
			\tabletypesize{\tiny}
			\tablehead{
				Line & Centroid (keV) & \multicolumn{11}{c}{Flux (10$^{-5}$\,photons\,cm$^{-2}$\,s$^{-1}$)} \\
				& & 
				2007 Jan. & 2008 Jan. & 2009 Jan. & 2009 Dec. & 2010 Dec. & 2011 Dec. & 2012 Dec. & 2014 Nov. & 2015 Nov. & 2016 Nov. & 2017 Oct. & 2019 Nov. 
			}
			\startdata
			N He$\alpha$ r &
			$0.4307$ &
			$22.34^{+3.65}_{-3.53}$ &
			$29.79^{+5.28}_{-3.98}$ &
			$17.63^{+3.26}_{-4.44}$ &
			$16.03^{+3.04}_{-2.24}$ &
			$21.09^{+6.59}_{-7.37}$ &
			$22.97^{+4.67}_{-9.41}$ &
			$13.61^{+3.21}_{-9.93}$ &
			$13.67^{+4.53}_{-4.91}$ &
			$5.19^{+2.85}_{-3.20}$ &
			$8.77^{+5.27}_{-5.06}$ &
			$15.21^{+6.93}_{-8.01}$ &
			$7.29^{+0.86}_{-0.19}$ \\
			N Ly$\alpha$ &
			$0.5002$ &
			$51.91^{+2.42}_{-2.24}$ &
			$57.09^{+2.22}_{-2.30}$ &
			$60.21^{+2.15}_{-1.94}$ &
			$61.41^{+3.07}_{-2.27}$ &
			$59.43^{+1.38}_{-1.20}$ &
			$61.61^{+1.68}_{-4.09}$ &
			$51.57^{+2.00}_{-2.89}$ &
			$48.67^{+1.54}_{-2.10}$ &
			$41.39^{+1.47}_{-3.50}$ &
			$38.80^{+2.27}_{-1.76}$ &
			$29.83^{+2.66}_{-2.23}$ &
			$39.45^{+1.33}_{-1.03}$ \\
			O He$\alpha$ f &
			$0.5611$ &
			$12.64^{+1.51}_{-2.18}$ &
			$15.06^{+2.35}_{-2.89}$ &
			$18.07^{+1.08}_{-1.41}$ &
			$13.65^{+2.08}_{-1.73}$ &
			$16.69^{+3.02}_{-0.99}$ &
			$14.08^{+2.31}_{-2.92}$ &
			$14.00^{+2.79}_{-1.22}$ &
			$7.61^{+2.02}_{-1.81}$ &
			$1.75^{+2.70}_{-1.31}$ &
			$5.69^{+1.83}_{-1.32}$ &
			$7.57^{+2.50}_{-2.56}$ &
			$<0.22$\\
			O He$\alpha$ i &
			$0.5686$ &
			$1.26^{+3.05}_{-0.93}$ &
			$4.28^{+2.45}_{-1.68}$ &
			$5.81^{+1.94}_{-2.89}$ &
			$3.66^{+1.14}_{-0.59}$ &
			$0.32^{+2.69}_{-0.24}$ &
			$4.23^{+1.15}_{-2.62}$ &
			$<3.28$ &
			$6.95^{+2.48}_{-1.55}$ &
			$5.60^{+2.85}_{-1.90}$ &
			$1.42^{+1.50}_{-1.07}$ &
			$3.41^{+1.60}_{-2.16}$ &
			$5.19^{+1.66}_{-0.76}$\\
			O He$\alpha$ r &
			$0.5740$ &
			$14.91^{+1.87}_{-1.89}$ &
			$18.03^{+3.41}_{-3.19}$ &
			$14.32^{+2.69}_{-1.72}$ &
			$14.09^{+1.36}_{-2.13}$ &
			$14.26^{+2.30}_{-3.18}$ &
			$12.32^{+4.16}_{-1.47}$ &
			$11.09^{+3.00}_{-1.86}$ &
			$13.88^{+2.86}_{-3.37}$ &
			$8.87^{+1.73}_{-1.05}$ &
			$10.66^{+2.15}_{-2.39}$ &
			$7.93^{+2.16}_{-1.32}$ &
			$4.73^{+0.42}_{-0.44}$\\
			O Ly$\alpha$ &
			$0.6537$ &
			$34.47^{+1.03}_{-0.80}$ &
			$43.24^{+2.20}_{-2.47}$ &
			$45.63^{+0.81}_{-1.85}$ &
			$49.01^{+2.90}_{-1.37}$ &
			$52.22^{+0.50}_{-1.48}$ &
			$51.07^{+1.37}_{-2.27}$ &
			$48.57^{+1.15}_{-0.41}$ &
			$37.28^{+0.78}_{-1.02}$ &
			$35.75^{+1.35}_{-1.18}$ &
			$37.22^{+0.92}_{-1.47}$ &
			$28.93^{+1.47}_{-1.39}$ &
			$25.87^{+0.30}_{-0.24}$\\
			O He$\beta$ &
			$0.6656$ &
			$3.10^{+0.58}_{-0.86}$ &
			$2.13^{+0.61}_{-0.35}$ &
			$3.51^{+0.63}_{-0.90}$ &
			$2.35^{+0.80}_{-0.92}$ &
			$0.15^{+0.88}_{-0.13}$ &
			$3.93^{+1.10}_{-1.13}$ &
			$1.66^{+0.33}_{-0.72}$ &
			$3.15^{+0.71}_{-0.73}$ &
			$3.63^{+1.17}_{-1.00}$ &
			$3.55^{+1.21}_{-0.82}$ &
			$2.66^{+0.52}_{-0.65}$ &
			$3.24^{+0.01}_{-0.22}$\\
			Fe XVII &
			$0.7252$ &
			$4.69^{+0.84}_{-1.16}$ &
			$8.73^{+1.61}_{-1.50}$ &
			$8.76^{+2.07}_{-1.58}$ &
			$11.19^{+0.99}_{-1.31}$ &
			$13.37^{+0.89}_{-1.76}$ &
			$13.21^{+2.05}_{-1.50}$ &
			$13.15^{+1.02}_{-1.64}$ &
			$7.58^{+1.00}_{-1.71}$ &
			$9.54^{+1.50}_{-2.10}$ &
			$4.05^{+1.80}_{-1.66}$ &
			$9.58^{+1.37}_{-1.52}$ &
			$5.89^{+0.28}_{-0.11}$ \\
			Fe XVII &
			$0.7271$ &
			$12.80^{+0.81}_{-0.92}$ &
			$17.12^{+1.06}_{-1.36}$ &
			$20.03^{+2.48}_{-1.82}$ &
			$20.53^{+0.65}_{-1.76}$ &
			$21.43^{+2.19}_{-2.12}$ &
			$21.87^{+1.72}_{-1.61}$ &
			$25.81^{+1.70}_{-2.25}$ &
			$22.07^{+1.55}_{-1.37}$ &
			$19.06\pm2.09$ &
			$19.16^{+1.75}_{-1.97}$ &
			$13.40^{+1.86}_{-1.36}$ &
			$11.08^{+2.85}_{-2.11}$\\
			Fe XVII &
			$0.7389$ &
			$7.31^{+0.64}_{-0.83}$ &
			$7.81^{+0.55}_{-0.70}$ &
			$10.65^{+0.45}_{-1.20}$ &
			$10.76^{+0.88}_{-0.50}$ &
			$14.09^{+1.03}_{-1.07}$ &
			$13.51^{+1.13}_{-0.72}$ &
			$11.49^{+1.15}_{-0.56}$ &
			$10.32^{+0.80}_{-1.04}$ &
			$11.30^{+0.33}_{-0.56}$ &
			$10.81^{+0.89}_{-1.13}$ &
			$10.25^{+0.56}_{-0.95}$ &
			$5.37^{+0.97}_{-1.30}$\\
			Fe XVIII &
			$0.7715$ &
			$2.75^{+0.49}_{-0.82}$ &
			$5.08^{+0.55}_{-0.44}$ &
			$4.76\pm0.49$ &
			$5.63^{+0.91}_{-1.31}$ &
			$8.73^{+0.60}_{-1.26}$ &
			$7.08^{+0.38}_{-0.92}$ &
			$5.90^{+0.48}_{-0.71}$ &
			$9.49^{+0.72}_{-1.19}$ &
			$8.43^{+0.61}_{-0.71}$ &
			$6.72^{+0.74}_{-0.75}$ &
			$7.84^{+0.87}_{-0.67}$ &
			$4.12^{+0.09}_{-0.16}$\\
			O Ly$\beta$ &
			$0.7746$\tablenotemark{\tiny a} &
			$6.88^{+0.68}_{-0.42}$ &
			$7.76^{+0.37}_{-0.43}$ &
			$10.07^{+0.63}_{-0.56}$ &
			$10.20^{+0.42}_{-1.07}$ &
			$11.49^{+1.28}_{-0.76}$ &
			$12.37^{+0.78}_{-0.57}$ &
			$12.09^{+0.58}_{-0.69}$ &
			$9.31^{+0.92}_{-0.83}$ &
			$7.14^{+0.78}_{-1.07}$ &
			$8.03^{+1.14}_{-0.66}$ &
			$7.84^{+0.52}_{-0.78}$ &
			$7.05^{+0.73}_{-0.94}$\\
			Fe XVII &
			$0.8124$ &
			$5.90^{+0.74}_{-0.55}$ &
			$9.10^{+0.54}_{-0.55}$ &
			$8.56^{+0.83}_{-0.78}$ &
			$10.95^{+1.13}_{-0.39}$ &
			$11.56^{+0.24}_{-1.03}$ &
			$13.49^{+1.29}_{-1.02}$ &
			$10.80^{+0.71}_{-0.82}$ &
			$9.30^{+0.37}_{-0.87}$ &
			$8.44^{+1.11}_{-0.97}$ &
			$10.19^{+0.88}_{-0.98}$ &
			$6.58^{+0.59}_{-0.62}$ &
			$7.28^{+0.20}_{-0.03}$\\
			O Ly$\gamma$ &
			$0.8170$ &
			$2.08^{+0.38}_{-0.75}$ &
			$3.44^{+0.45}_{-0.49}$ &
			$4.19^{+0.48}_{-0.42}$ &
			$5.92^{+0.16}_{-1.29}$ &
			$5.42^{+1.10}_{-1.13}$ &
			$5.37^{+0.43}_{-1.11}$ &
			$5.30^{+0.45}_{-1.06}$ &
			$4.70^{+0.79}_{-0.50}$ &
			$4.46^{+0.48}_{-0.59}$ &
			$2.99^{+0.91}_{-1.16}$ &
			$4.25^{+1.16}_{-0.87}$ &
			$1.14^{+0.14}_{-0.15}$\\
			Fe XVII &
			$0.8258$ &
			$15.05^{+0.38}_{-0.61}$ &
			$17.86^{+0.48}_{-0.53}$ &
			$22.98^{+0.86}_{-0.61}$ &
			$26.15^{+0.58}_{-0.82}$ &
			$30.00^{+0.83}_{-0.60}$ &
			$30.01^{+0.61}_{-0.95}$ &
			$29.76^{+1.30}_{-0.66}$ &
			$25.82^{+0.62}_{-0.90}$ &
			$26.42^{+0.51}_{-0.64}$ &
			$23.95^{+1.42}_{-0.80}$ &
			$20.90^{+1.18}_{-1.07}$ &
			$17.79^{+0.21}_{-0.29}$\\
			Fe XVIII &
			$0.8531$ &
			$2.12^{+0.46}_{-0.44}$ &
			$3.47^{+0.27}_{-0.40}$ &
			$3.66^{+0.38}_{-0.43}$ &
			$3.43^{+0.36}_{-0.61}$ &
			$4.65^{+0.68}_{-0.67}$ &
			$4.69^{+0.65}_{-0.56}$ &
			$4.62^{+0.64}_{-0.44}$ &
			$4.86^{+0.44}_{-0.46}$ &
			$5.60^{+0.85}_{-0.44}$ &
			$6.60^{+0.86}_{-0.69}$ &
			$3.52^{+0.64}_{-0.54}$ &
			$3.19^{+0.15}_{-0.07}$\\
			Fe XVIII &
			$0.8626$ &
			$1.93^{+0.42}_{-0.27}$ &
			$3.44^{+0.63}_{-0.47}$ &
			$3.14^{+0.30}_{-0.29}$ &
			$5.36^{+0.30}_{-0.51}$ &
			$5.87^{+0.72}_{-0.74}$ &
			$5.62^{+1.10}_{-0.43}$ &
			$6.47^{+0.52}_{-0.75}$ &
			$5.86^{+0.70}_{-0.71}$ &
			$4.96^{+0.75}_{-0.45}$ &
			$6.01^{+0.74}_{-0.77}$ &
			$3.85^{+0.80}_{-0.60}$ &
			$3.94^{+0.13}_{-0.11}$\\
			Fe XVIII &
			$0.8726$ &
			$4.56^{+0.35}_{-0.36}$ &
			$6.06^{+0.57}_{-0.78}$ &
			$8.27^{+0.20}_{-0.59}$ &
			$9.54^{+0.37}_{-0.15}$ &
			$11.72^{+0.50}_{-0.41}$ &
			$13.89^{+1.08}_{-0.72}$ &
			$12.97^{+0.71}_{-0.81}$ &
			$13.12^{+0.65}_{-0.98}$ &
			$11.92^{+0.63}_{-0.85}$ &
			$9.65^{+0.54}_{-0.69}$ &
			$10.11^{+0.58}_{-0.49}$ &
			$8.14^{+0.10}_{-0.08}$\\
			Fe XVII &
			$0.8968$ &
			$1.95^{+0.22}_{-0.60}$ &
			$2.18^{+0.30}_{-0.44}$ &
			$2.38^{+0.28}_{-0.17}$ &
			$3.96^{+0.21}_{-0.28}$ &
			$4.12^{+0.88}_{-0.39}$ &
			$5.80^{+0.93}_{-0.52}$ &
			$4.44^{+0.66}_{-0.25}$ &
			$5.17^{+0.64}_{-0.54}$ &
			$4.90^{+0.72}_{-0.71}$ &
			$4.97^{+0.96}_{-0.65}$ &
			$5.62^{+0.31}_{-0.43}$ &
			$3.79^{+0.44}_{-0.24}$\\
			Ne He$\alpha$ f &
			$0.9051$ &
			$7.05^{+0.58}_{-0.24}$ &
			$9.86^{+0.51}_{-0.35}$ &
			$9.36^{+0.60}_{-0.93}$ &
			$10.92^{+0.41}_{-1.03}$ &
			$13.63^{+0.91}_{-1.02}$ &
			$12.50^{+0.85}_{-0.82}$ &
			$13.49^{+0.79}_{-1.10}$ &
			$12.31^{+0.98}_{-0.67}$ &
			$10.27^{+0.96}_{-1.43}$ &
			$9.16^{+0.85}_{-0.71}$ &
			$7.56^{+1.35}_{-0.87}$ &
			$7.99^{+1.02}_{-0.98}$\\
			Fe XIX &
			$0.9172$\tablenotemark{\tiny b} &
			$5.15^{+0.70}_{-0.49}$ &
			$5.61^{+1.03}_{-0.60}$ &
			$9.22^{+0.82}_{-0.63}$ &
			$7.42^{+1.06}_{-0.99}$ &
			$8.28^{+0.92}_{-0.22}$ &
			$9.66^{+1.02}_{-0.45}$ &
			$11.09^{+0.34}_{-0.76}$ &
			$10.65^{+0.86}_{-0.69}$ &
			$11.26^{+0.83}_{-0.88}$ &
			$10.52^{+0.86}_{-0.78}$ &
			$8.44^{+0.78}_{-1.19}$ &
			$7.20^{+0.27}_{-0.35}$\\
			Ne He$\alpha$ r &
			$0.9220$ &
			$12.45^{+0.86}_{-0.75}$ &
			$13.99^{+0.81}_{-1.24}$ &
			$15.04^{+0.53}_{-1.11}$ &
			$16.59^{+0.62}_{-0.47}$ &
			$19.94^{+1.09}_{-1.71}$ &
			$19.70^{+0.46}_{-1.56}$ &
			$19.05^{+0.97}_{-0.82}$ &
			$18.95^{+1.00}_{-1.20}$ &
			$13.52^{+0.84}_{-0.72}$ &
			$14.85^{+1.10}_{-1.09}$ &
			$14.30^{+1.34}_{-1.02}$ &
			$13.87^{+2.34}_{-3.02}$\\
			Fe XX &
			$0.9652$ &
			$1.81^{+0.36}_{-0.50}$ &
			$3.39^{+0.45}_{-0.34}$ &
			$3.48^{+0.16}_{-0.43}$ &
			$5.48^{+0.50}_{-0.61}$ &
			$6.19^{+0.64}_{-0.72}$ &
			$7.61^{+0.58}_{-0.78}$ &
			$8.59^{+0.74}_{-0.87}$ &
			$7.85^{+0.71}_{-0.69}$ &
			$9.49^{+0.77}_{-0.60}$ &
			$9.12^{+0.47}_{-0.43}$ &
			$8.01^{+0.78}_{-0.55}$ &
			$7.37^{+0.58}_{-0.89}$\\
			Fe XXI &
			$1.0004$ &
			$1.17\pm0.30$ &
			$1.79^{+0.42}_{-0.49}$ &
			$0.92^{+0.30}_{-0.15}$ &
			$1.49^{+0.44}_{-0.57}$ &
			$3.48^{+0.66}_{-0.38}$ &
			$2.75^{+0.70}_{-0.56}$ &
			$2.61^{+0.15}_{-1.35}$ &
			$3.62^{+0.62}_{-0.44}$ &
			$3.54^{+0.59}_{-0.80}$ &
			$3.10^{+0.60}_{-0.38}$ &
			$4.14^{+0.70}_{-0.82}$ &
			$2.78^{+0.45}_{-0.44}$\\
			Fe XXI &
			$1.0093$\tablenotemark{\tiny c} &
			$2.35^{+0.40}_{-0.47}$ &
			$2.68^{+0.31}_{-0.36}$ &
			$3.55^{+0.38}_{-0.30}$ &
			$5.48^{+0.70}_{-0.58}$ &
			$4.66^{+0.55}_{-1.17}$ &
			$7.35^{+0.36}_{-1.18}$ &
			$7.44^{+0.44}_{-0.30}$ &
			$8.40^{+0.55}_{-0.67}$ &
			$7.63^{+0.99}_{-0.67}$ &
			$5.51^{+0.65}_{-0.72}$ &
			$5.80^{+0.89}_{-1.03}$ &
			$7.22^{+0.39}_{-0.21}$\\
			Ne Ly$\alpha$ &
			$1.0220$ &
			$12.18^{+0.24}_{-0.51}$ &
			$15.70^{+0.84}_{-0.94}$ &
			$19.75^{+0.31}_{-0.33}$ &
			$22.03^{+0.77}_{-0.31}$ &
			$26.47^{+0.20}_{-0.74}$ &
			$26.46^{+1.26}_{-0.52}$ &
			$26.69^{+1.34}_{-0.27}$ &
			$26.24^{+1.25}_{-1.37}$ &
			$24.40^{+1.43}_{-0.95}$ &
			$23.45^{+0.52}_{-0.74}$ &
			$22.96^{+1.34}_{-1.19}$ &
			$17.75^{+0.87}_{-1.20}$\\
			Ne He$\beta$ &
			$1.0740$ &
			$2.61^{+0.30}_{-0.34}$ &
			$2.75^{+0.32}_{-0.20}$ &
			$2.99^{+0.46}_{-0.21}$ &
			$2.95^{+0.58}_{-0.43}$ &
			$3.65^{+0.43}_{-0.57}$ &
			$4.74^{+0.58}_{-0.12}$ &
			$5.17^{+0.77}_{-0.07}$ &
			$4.94^{+0.51}_{-0.33}$ &
			$5.93^{+0.82}_{-0.35}$ &
			$4.58^{+0.97}_{-0.41}$ &
			$3.60^{+0.59}_{-0.70}$ &
			$4.62^{+0.23}_{-0.15}$\\
			Fe XXII &
			$1.0534$\tablenotemark{\tiny d} &
			$0.59^{+0.32}_{-0.19}$ &
			$1.60^{+0.37}_{-0.31}$ &
			$1.91^{+0.53}_{-0.36}$ &
			$3.26^{+0.46}_{-0.51}$ &
			$3.26^{+0.26}_{-0.37}$ &
			$4.57^{+0.33}_{-0.57}$ &
			$5.53^{+0.05}_{-1.10}$ &
			$7.16^{+0.65}_{-0.67}$ &
			$5.13^{+0.71}_{-0.48}$ &
			$6.12^{+0.44}_{-0.72}$ &
			$7.08^{+0.84}_{-0.58}$ &
			$7.23^{+0.72}_{-0.76}$\\
			Fe XXIII &
			$1.1291$\tablenotemark{\tiny e} &
			$1.39\pm0.18$ &
			$1.75^{+0.26}_{-0.31}$ &
			$1.89^{+0.40}_{-0.28}$ &
			$1.85^{+0.23}_{-0.44}$ &
			$2.83^{+0.34}_{-0.54}$ &
			$2.82^{+0.28}_{-0.67}$ &
			$4.18^{+0.12}_{-0.69}$ &
			$4.11^{+0.46}_{-0.47}$ &
			$5.44^{+0.67}_{-0.37}$ &
			$3.66^{+0.58}_{-0.46}$ &
			$3.37^{+0.62}_{-0.55}$ &
			$3.82^{+0.27}_{-0.49}$\\
			Ne Ly$\beta$ &
			$1.2110$ &
			$1.48^{+0.23}_{-0.30}$ &
			$1.88^{+0.22}_{-0.19}$ &
			$2.16^{+0.37}_{-0.32}$ &
			$2.51^{+0.18}_{-0.35}$ &
			$3.07^{+0.37}_{-0.52}$ &
			$3.18^{+0.48}_{-0.25}$ &
			$3.07^{+0.61}_{-0.37}$ &
			$2.88^{+0.43}_{-0.39}$ &
			$2.62^{+0.43}_{-0.24}$ &
			$2.94^{+0.39}_{-0.55}$ &
			$2.34^{+0.41}_{-0.35}$ &
			$1.33^{+0.08}_{-0.11}$\\
			Mg He$\alpha$ f &
			$1.3311$ &
			$1.92^{+0.17}_{-0.21}$ &
			$2.07^{+0.38}_{-0.37}$ &
			$3.09^{+0.25}_{-0.28}$ &
			$3.23^{+0.46}_{-0.34}$ &
			$3.98^{+0.17}_{-0.14}$ &
			$3.54^{+0.65}_{-0.52}$ &
			$3.61^{+0.23}_{-0.19}$ &
			$2.02^{+0.40}_{-0.25}$ &
			$2.87^{+0.58}_{-0.55}$ &
			$3.19^{+0.33}_{-0.48}$ &
			$2.68^{+0.54}_{-0.47}$ &
			$2.65^{+0.44}_{-0.59}$\\
			Mg He$\alpha$ r &
			$1.3523$ &
			$3.02^{+0.31}_{-0.29}$ &
			$3.75^{+0.54}_{-0.46}$ &
			$4.80^{+0.12}_{-0.21}$ &
			$5.24^{+0.28}_{-0.34}$ &
			$4.84^{+0.32}_{-0.38}$ &
			$7.07^{+0.52}_{-0.28}$ &
			$6.48^{+0.37}_{-0.49}$ &
			$6.54^{+0.63}_{-0.67}$ &
			$5.58^{+0.65}_{-0.44}$ &
			$5.34^{+0.41}_{-0.42}$ &
			$4.73^{+0.42}_{-0.33}$ &
			$3.36^{+1.44}_{-1.10}$\\
			Mg Ly$\alpha$ &
			$1.4726$ &
			$1.30^{+0.23}_{-0.31}$ &
			$1.98^{+0.29}_{-0.27}$ &
			$2.59^{+0.14}_{-0.26}$ &
			$2.87^{+0.22}_{-0.17}$ &
			$2.69^{+0.40}_{-0.36}$ &
			$3.32^{+0.50}_{-0.23}$ &
			$3.80^{+0.33}_{-0.22}$ &
			$4.09^{+0.44}_{-0.39}$ &
			$4.36^{+0.46}_{-0.63}$ &
			$3.59^{+0.51}_{-0.49}$ &
			$4.03^{+0.34}_{-0.32}$ &
			$3.46^{+0.13}_{-0.14}$\\
			Si He$\alpha$ f &
			$1.8395$ &
			$1.14^{+0.31}_{-0.53}$ &
			$0.79^{+0.48}_{-0.40}$ &
			$0.51^{+0.55}_{-0.21}$ &
			$1.68^{+0.44}_{-0.41}$ &
			$1.54^{+0.55}_{-0.38}$ &
			$2.18^{+0.66}_{-0.97}$ &
			$1.91^{+0.36}_{-0.48}$ &
			$0.31^{+0.57}_{-0.24}$ &
			$2.54^{+0.69}_{-1.10}$ &
			$1.62^{+0.80}_{-1.07}$ &
			$1.11^{+0.73}_{-0.61}$ &
			$<0.06$\\
			Si He$\alpha$ r &
			$1.8650$ &
			$2.55^{+0.49}_{-0.45}$ &
			$3.27^{+0.44}_{-0.53}$ &
			$4.39^{+0.50}_{-0.84}$ &
			$4.93^{+0.52}_{-0.82}$ &
			$5.82^{+0.45}_{-1.11}$ &
			$4.98^{+0.96}_{-0.84}$ &
			$6.19^{+0.97}_{-0.70}$ &
			$6.97^{+0.48}_{-0.56}$ &
			$4.24^{+1.41}_{-0.96}$ &
			$5.00^{+1.05}_{-1.31}$ &
			$5.81^{+0.46}_{-0.69}$ &
			$6.19^{+0.26}_{-0.49}$\\
			Si Ly$\alpha$ &
			$2.0061$ &
			$1.61^{+0.74}_{-0.73}$ &
			$1.06^{+0.55}_{-0.72}$ &
			$0.11^{+1.15}_{0.09}$ &
			$<1.05$ &
			$1.72^{+0.82}_{-0.95}$ &
			$2.16^{+0.93}_{-0.96}$ &
			$0.04^{+0.65}_{-0.02}$ &
			$1.93^{+1.28}_{-1.01}$ &
			$3.42^{+1.30}_{-0.81}$ &
			$3.66^{+1.01}_{-0.66}$ &
			$1.98^{+1.45}_{-1.32}$ &
			$3.56^{+0.70}_{-0.41}$\\
			\hline
			$v_{\rm 87A}$ (km\,s$^{-1}$) &
			&
			$246^{+10}_{-13}$ &
			$283^{+6}_{-9}$ &
			$226^{+20}_{-7}$ &
			$200^{+8}_{-9}$ &
			$230^{+21}_{-7}$ &
			$220^{+7}_{-19}$ &
			$254^{+13}_{-9}$ &
			$283^{+13}_{-8}$ &
			$360^{+14}_{-18}$ &
			$255^{+28}_{-31}$ &
			$134^{+27}_{-20}$ &
			$59^{+12}_{-14}$\\
			$\sigma_6$ (eV) &
			&
			$92^{+53}_{-26}$ &
			$66^{+35}_{-25}$ &
			$107^{+68}_{-25}$ &
			$43^{+22}_{-16}$ &
			$84\pm19$ &
			$134^{+58}_{-31}$ &
			$95^{+23}_{-16}$ &
			$155^{+77}_{-34}$ &
			$137^{+67}_{-34}$ &
			$101^{+36}_{-29}$ &
			$85^{+22}_{-19}$ &
			$5\pm1$\\
			$\alpha$ &
			&
			$2.25^{+0.27}_{-0.14}$ &
			$2.09^{+0.27}_{-0.20}$ &
			$2.37^{+0.33}_{-0.13}$ &
			$1.90^{+0.24}_{-0.22}$ &
			$2.15\pm0.12$ &
			$2.01^{+0.22}_{-0.12}$ &
			$2.22^{+0.16}_{-0.07}$ &
			$2.47^{+0.24}_{-0.11}$ &
			$2.01^{+0.23}_{-0.13}$ &
			$2.04^{+0.17}_{-0.15}$ &
			$2.03^{+0.12}_{-0.13}$&
			$0.48^{+0.04}_{-0.01}$\\
			\enddata
			\tablecomments{Errors are for 1-$\sigma$ confidence range.}
			\tablenotetext{a}{Could also be Fe XVIII line at 0.7747 keV}
			\tablenotetext{b}{Could also be Fe XXI line at 0.9179 keV}
			\tablenotetext{c}{Could also be Fe XVII line at 1.0108 keV}
			\tablenotetext{d}{Could also be Fe XXIII line at 1.0565 keV}
			\tablenotetext{e}{Could also be Ne He$\gamma$ line at 1.1270 keV}
		\end{deluxetable*}
	\end{longrotatetable}
	
	\clearpage

		\begin{longrotatetable}
		\begin{deluxetable*}{lcccccccccccc}
			\tablecaption{Spectral Fitting Results of the Two-temperature {\tt vnei} Model\label{tab:2-T nei_nH}}
			\tablenum{3}
			\tabletypesize{\tiny}
			\tablehead{
				& 2007 Jan. & 2008 Jan. & 2009 Jan. & 2009 Dec. & 2010 Dec. & 2011 Dec. & 2012 Dec. & 2014 Nov. & 2015 Nov. & 2016 Nov. & 2017 Oct. & 2019 Nov.
			}
			\startdata
			\multicolumn{13}{l}{Low-Temperature Plasma} \\
			$kT_{\rm LT}$ (keV) & $0.56\pm0.01$ & $0.59^{+0.02}_{-0.01}$ & $0.55\pm0.01$ & $0.57\pm0.01$ & $0.60\pm0.01$ & $0.61\pm0.01$ & $0.62\pm0.01$ & $0.67\pm0.01$ & $0.65^{+0.02}_{-0.01}$ & $0.69\pm0.01$ & $0.66^{+0.02}_{-0.01}$ & $0.71\pm0.03$\\
			$\tau_{\rm LT}$ ($10^{10}$\,cm$^{-3}$\,s) & $3.93^{+0.14}_{-0.09}$ & $3.98^{+0.05}_{-0.04}$ & $8.51^{+0.23}_{-0.17}$ & $6.75^{+0.20}_{-0.18}$ & $8.18^{+0.27}_{-0.17}$ & $7.96^{+0.32}_{-0.18}$ & $6.98^{+0.12}_{-0.11}$ & $8.18^{+0.25}_{-0.23}$ & $8.03^{+0.71}_{-0.39}$ & $7.74^{+0.30}_{-0.27}$ & $7.31^{+0.16}_{-0.13}$ & $4.73^{+0.28}_{-0.18}$\\
			Norm$_{\rm LT}$ ($10^{-3}$\,cm$^{-5}$) & $3.48^{+0.23}_{-0.14}$ &$4.14^{+0.10}_{-0.16}$ & $6.96^{+0.29}_{-0.21}$ & $6.66\pm0.27$ & $7.20^{+0.19}_{-0.30}$ & $7.10^{+0.18}_{-0.25}$ & $6.98^{+0.20}_{-0.19}$ & $6.81^{+0.19}_{-0.22}$ & $6.57^{+0.31}_{-0.26}$ & $6.03^{+0.21}_{-0.24}$ & $5.72^{+0.19}_{-0.25}$ & $4.05^{+0.26}_{-0.22}$\\
			\hline
			\multicolumn{13}{l}{High-Temperature Plasma} \\
			$kT_{\rm HT}$ (keV) & $2.27^{+0.14}_{-0.12}$ & $2.28^{+0.11}_{-0.09}$ & $2.33^{+0.14}_{-0.09}$ & $2.25^{+0.08}_{-0.07}$ & $2.35\pm0.10$ & $2.42\pm0.07$ & $2.34^{+0.08}_{-0.07}$ & $2.53^{+0.09}_{-0.08}$ & $2.51^{+0.10}_{-0.06}$ & $2.66^{+0.13}_{-0.09}$ & $2.52^{+0.13}_{-0.09}$ & $2.75^{+0.08}_{-0.15}$\\
			$\tau_{\rm HT}$ ($10^{10}$\,cm$^{-3}$\,s) & $4.58^{+0.46}_{-0.31}$ & $5.16^{+0.41}_{-0.37}$ & $5.33^{+0.40}_{-0.29}$ & $5.21^{+0.29}_{-0.26}$ & $6.73^{+0.78}_{-0.48}$ & $6.11^{+0.54}_{-0.31}$ & $6.18^{+0.40}_{-0.38}$ & $6.88^{+0.31}_{-0.29}$ & $5.88^{+0.48}_{-0.27}$ & $6.69^{+0.36}_{-0.32}$ & $6.78^{+0.50}_{-0.40}$ & $6.26^{+0.84}_{-0.53}$\\
			Norm$_{\rm HT}$ ($10^{-3}$\,cm$^{-5}$) & $1.12^{+0.07}_{-0.08}$ & $1.47^{+0.07}_{-0.08}$ & $1.73^{+0.08}_{-0.14}$ & $2.20\pm0.11$ & $2.31^{+0.09}_{-0.10}$ & $2.68\pm0.08$ & $3.19\pm0.11$ & $3.19^{+0.11}_{-0.12}$ & $3.54^{+0.14}_{-0.18}$ & $3.42^{+0.12}_{-0.16}$ & $3.93^{+0.13}_{-0.17}$ & $3.89^{+0.20}_{-0.11}$\\
			\hline
			\multicolumn{13}{l}{Abundance} \\
			N & $0.69^{+0.04}_{-0.06}$ & $0.71^{+0.07}_{-0.04}$ & $1.31^{+0.21}_{-0.12}$ & $0.94^{+0.12}_{-0.08}$ & $1.50^{+0.25}_{-0.18}$ & $1.44^{+0.21}_{-0.12}$ & $0.88^{+0.09}_{-0.12}$ & $1.54^{+0.23}_{-0.19}$ & $1.10^{+0.34}_{-0.13}$ & $1.27^{+0.20}_{-0.16}$ & $0.73^{+0.17}_{-0.12}$ & $0.56^{+0.13}_{-0.11}$\\
			O & $0.08\pm0.01$ & $0.08\pm0.01$ & $0.08\pm0.01$ & $0.07\pm0.01$ & $0.09\pm0.01$ & $0.08\pm0.01$ & $0.07\pm0.01$ & $0.07\pm0.01$ & $0.06\pm0.01$ & $0.07\pm0.01$ & $0.06\pm0.01$ & $0.05\pm0.01$\\
			Ne & $0.31\pm0.02$ & $0.30\pm0.02$ & $0.29\pm0.01$ & $0.27\pm0.01$ & $0.32^{+0.03}_{-0.02}$ & $0.30^{+0.02}_{-0.01}$ & $0.28\pm0.02$ & $0.29\pm0.02$ & $0.24^{+0.02}_{-0.01}$ & $0.26^{+0.02}_{-0.01}$ & $0.24\pm0.02$ & $0.23^{+0.04}_{-0.03}$\\
			Mg & $0.39\pm0.03$ & $0.38^{+0.03}_{-0.02}$ & $0.37\pm0.02$ & $0.37\pm0.02$ & $0.41^{+0.04}_{-0.02}$ & $0.43^{+0.03}_{-0.02}$ & $0.39^{+0.03}_{-0.02}$ & $0.38^{+0.03}_{-0.02}$ & $0.33\pm0.03$ & $0.36^{+0.03}_{-0.02}$ & $0.33\pm0.03$ & $0.30\pm0.04$\\
			Si & $0.59^{+0.05}_{-0.07}$ & $0.61^{+0.06}_{-0.05}$ & $0.63^{+0.05}_{-0.04}$ & $0.61\pm0.04$ & $0.67^{+0.06}_{-0.04}$ & $0.66\pm0.04$ & $0.67^{+0.05}_{-0.04}$ & $0.63^{+0.04}_{-0.03}$ & $0.59\pm0.04$ & $0.63^{+0.04}_{-0.03}$ & $0.58^{+0.05}_{-0.06}$ & $0.61^{+0.06}_{-0.07}$\\
			S & $0.65^{+0.11}_{-0.10}$ & $0.52^{+0.08}_{-0.09}$ & $0.61\pm0.09$ & $0.48^{+0.07}_{-0.06}$ & $0.58\pm0.09$ & $0.56^{+0.08}_{-0.06}$ & $0.55\pm0.07$ & $0.57\pm0.07$ & $0.48^{+0.07}_{-0.09}$ & $0.46^{+0.08}_{-0.07}$ & $0.44^{+0.10}_{-0.09}$ & $0.56^{+0.11}_{-0.09}$\\
			Fe & $0.26^{+0.01}_{-0.02}$ & $0.29^{+0.02}_{-0.01}$ & $0.23\pm0.01$ & $0.26\pm0.01$ & $0.28^{+0.02}_{-0.01}$ & $0.29\pm0.01$ & $0.29\pm0.01$ & $0.28\pm0.01$ & $0.26\pm0.01$ & $0.27\pm0.01$ & $0.26^{+0.02}_{-0.01}$ & $0.26\pm0.02$\\
			\hline
			$\chi^2_{\rm r}$ & 1.35 & 1.33 & 1.34  & 1.41  & 1.34 & 1.35  & 1.31  & 1.32  & 1.33  & 1.36  & 1.21  & 1.21 \\
			($\chi^2/$dof) & (2737/2028) & (2891/2175) & (3249/2433) & (3353/2371) &  (2650/1974) & (3164/2347) & (2907/2216) & (3118/2363) & (2872/2153) & (3106/2283) & (2731/2263) & (1494/1234)\\
			\enddata
			\tablecomments{Errors are for 90\% confidence range.}
		\end{deluxetable*}
	\end{longrotatetable}

	\clearpage

		\begin{longrotatetable}
		\begin{deluxetable*}{lcccccccccccc}
			\tablecaption{Spectral Fitting Results of the Two-temperature {\tt vpshock} Model\label{tab:2-T pshock_nH}}
			\tablenum{4}
			\tabletypesize{\tiny}
			\tablehead{
				& 2007 Jan. & 2008 Jan. & 2009 Jan. & 2009 Dec. & 2010 Dec. & 2011 Dec. & 2012 Dec. & 2014 Nov. & 2015 Nov. & 2016 Nov. & 2017 Oct. & 2019 Nov.
			}
			\startdata
			\multicolumn{13}{l}{Low-Temperature Plasma} \\
			$kT_{\rm LT}$ (keV) & $0.55^{+0.02}_{-0.01}$ & $0.58\pm0.01$ & $0.60^{+0.01}_{-0.02}$ & $0.60\pm0.01$ & $0.62\pm0.01$ & $0.63\pm0.01$& $0.63\pm0.01$ & $0.68\pm0.01$ & $0.66\pm0.01$ & $0.69\pm0.01$ & $0.63\pm0.02$ & $0.63\pm0.03$\\
			$\tau_{\rm u,LT}$ ($10^{11}$\,cm$^{-3}$\,s) & $2.85^{+0.22}_{-0.13}$ & $3.26^{+0.15}_{-0.17}$ & $4.24^{+0.40}_{-0.23}$ & $4.30^{+0.20}_{-0.27}$ & $5.28^{+0.25}_{-0.33}$ & $6.02^{+0.08}_{-0.12}$ & $6.17^{+0.16}_{-0.24}$ & $7.34^{+0.58}_{-0.33}$ & $9.93^{+0.20}_{-0.10}$ & $6.75^{+0.37}_{-0.43}$ & $6.39^{+0.13}_{-0.19}$ & $8.64^{+0.26}_{-0.12}$\\
			Norm$_{\rm LT}$ ($10^{-3}$\,cm$^{-5}$) & $3.87^{+0.31}_{-0.25}$ & $4.45^{+0.42}_{-0.38}$ & $6.07^{+0.34}_{-0.25}$ & $6.12^{+0.21}_{-0.34}$ & $6.10^{+0.35}_{-0.32}$ & $5.98^{+0.40}_{-0.32}$ & $6.08^{+0.30}_{-0.41}$ & $5.60^{+0.41}_{-0.32}$ & $5.45^{+0.27}_{-0.37}$ & $4.62^{+0.35}_{-0.23}$ & $4.29^{+0.20}_{-0.31}$ & $3.03^{+0.40}_{-0.30}$\\
			\hline
			\multicolumn{13}{l}{High-Temperature Plasma} \\
			$kT_{\rm HT}$ (keV) & $2.52\pm0.21$ & $2.49^{+0.21}_{-0.14}$ & $2.54^{+0.17}_{-0.16}$ & $2.42^{+0.09}_{-0.11}$ & $2.34^{+0.12}_{-0.08}$ & $2.42^{+0.11}_{-0.08}$ & $2.33^{+0.07}_{-0.10}$ & $2.42^{+0.08}_{-0.09}$ & $2.44\pm0.09$ & $2.47\pm0.07$ & $2.33^{+0.09}_{-0.06}$ & $2.53^{+0.13}_{-0.07}$\\
			$\tau_{\rm u,HT}$ ($10^{11}$\,cm$^{-3}$\,s) & $0.96^{+0.17}_{-0.10}$ & $1.28^{+0.08}_{-0.09}$ & $1.32^{+0.13}_{-0.08}$ & $1.28^{+0.03}_{-0.04}$ & $2.19^{+0.05}_{-0.07}$ & $1.64\pm0.01$ & $1.67\pm0.01$ & $2.17^{+0.05}_{-0.03}$ & $1.72\pm0.01$ & $2.09^{+0.03}_{-0.04}$ & $1.72\pm0.01$ & $1.67^{+0.02}_{-0.01}$\\
			Norm$_{\rm HT}$ ($10^{-3}$\,cm$^{-5}$) & $0.97\pm0.12$ & $1.30^{+0.12}_{-0.13}$ & $1.48^{+0.14}_{-0.12}$ & $1.92^{+0.15}_{-0.11}$ & $2.26^{+0.08}_{-0.13}$ & $2.63^{+0.12}_{-0.15}$ & $3.14^{+0.18}_{-0.12}$ & $3.34^{+0.13}_{-0.14}$ & $3.61^{+0.17}_{-0.15}$ & $3.74\pm0.14$ & $4.44^{+0.17}_{-0.19}$ & $4.36^{+0.16}_{-0.24}$\\
			\hline
			\multicolumn{13}{l}{Abundance} \\
			N & $1.46^{+0.17}_{-0.14}$ & $1.73^{+0.16}_{-0.18}$ & $1.66^{+0.17}_{-0.16}$ &$1.54^{+0.12}_{-0.11}$ & $1.95^{+0.18}_{-0.22}$ & $1.79^{+0.15}_{-0.13}$ & $1.36^{+0.11}_{-0.13}$ & $1.58^{+0.15}_{-0.12}$ & $1.29^{+0.15}_{-0.12}$ & $1.30^{+0.14}_{-0.12}$ & $0.74^{+0.11}_{-0.06}$ & $1.05^{+0.16}_{-0.15}$\\
			O &  $0.14^{+0.02}_{-0.01}$ & $0.16\pm0.01$ & $0.17^{+0.01}_{-0.02}$ & $0.16\pm0.01$ & $0.21\pm0.02$ & $0.19\pm0.01$ & $0.17\pm0.01$ & $0.19\pm0.01$ & $0.16\pm0.01$ & $0.17\pm0.01$ & $0.11\pm0.01$ & $0.13\pm0.02$\\
			Ne & $0.41\pm0.03$ & $0.43^{+0.02}_{-0.03}$ & $0.42^{+0.03}_{-0.02}$ & $0.41\pm0.02$ & $0.54^{+0.03}_{-0.04}$ & $0.50^{+0.03}_{-0.02}$ & $0.47^{+0.02}_{-0.03}$ & $0.52\pm0.03$ & $0.45\pm0.03$ & $0.46\pm0.03$ & $0.34\pm0.03$ & $0.34^{+0.05}_{-0.03}$\\
			Mg & $0.46^{+0.04}_{-0.03}$ & $0.47\pm0.03$ & $0.47^{+0.04}_{-0.03}$ & $0.47\pm0.03$ & $0.59\pm0.04$ & $0.59^{+0.04}_{-0.03}$ & $0.53^{+0.04}_{-0.03}$ & $0.57^{+0.03}_{-0.04}$ & $0.49\pm0.04$ & $0.53^{+0.03}_{-0.04}$ & $0.42^{+0.05}_{-0.04}$ & $0.41^{+0.07}_{-0.05}$\\
			Si & $0.67^{+0.07}_{-0.05}$ & $0.72^{+0.04}_{-0.06}$ & $0.72^{+0.06}_{-0.04}$ & $0.72^{+0.05}_{-0.04}$ & $0.83^{+0.07}_{-0.06}$ & $0.79\pm0.05$ & $0.79^{+0.04}_{-0.05}$ & $0.77^{+0.04}_{-0.05}$ & $0.72^{+0.06}_{-0.05}$ & $0.77^{+0.07}_{-0.05}$ & $0.64^{+0.05}_{-0.06}$ & $0.72\pm0.09$\\
			S & $0.71^{+0.16}_{-0.12}$ & $0.54^{+0.08}_{-0.09}$ & $0.66\pm0.10$ & $0.53^{+0.09}_{-0.08}$ & $0.63^{+0.12}_{-0.11}$ & $0.62\pm0.08$ & $0.61^{+0.08}_{-0.07}$ & $0.63^{+0.09}_{-0.06}$ & $0.53\pm0.09$ & $0.51^{+0.10}_{-0.08}$ & $0.45\pm0.10$ & $0.62^{+0.13}_{-0.11}$\\
			Fe & $0.26\pm0.02$ & $0.30^{+0.03}_{-0.02}$ & $0.27^{+0.01}_{-0.02}$ & $0.30^{+0.02}_{-0.01}$ & $0.35\pm0.02$ & $0.36^{+0.02}_{-0.03}$ & $0.34^{+0.03}_{-0.02}$ & $0.36\pm0.02$ & $0.33\pm0.02$ & $0.37^{+0.02}_{-0.03}$ & $0.32^{+0.02}_{-0.01}$ & $0.35^{+0.04}_{-0.03}$\\
			\hline
			$\chi^2_{\rm r}$& 1.27 & 1.22 & 1.24 & 1.33 & 1.26 & 1.26 & 1.26 & 1.24 & 1.28 & 1.30 & 1.17 & 1.18\\
			($\chi^2/$dof)& (2582/2028) & (2647/2175) & (3010/2433) & (3150/2371) & (2483/1974) & (2955/2347) & (2797/2216) & (2932/2363) & (2746/2153) & (2976/2283) & (2641/2263) & (1450/1234)\\
			\enddata
			\tablecomments{Errors are for 90\% confidence range.}
		\end{deluxetable*}
	\end{longrotatetable}

	\clearpage
	
		\begin{deluxetable*}{lccccccc}
		\tablecaption{Fitting Results of the Fe K Line\label{tab:Fe K_AIC}}
		\tablenum{5}
		\tabletypesize{\normalsize}
		\tablehead{
			& Centroid (keV) & Flux (10$^{-6}$\,ph\,cm$^{-2}$\,s$^{-1}$) & EW (eV) & $\chi^2_{\rm r}$ ($\chi^2$/dof) & F-test p-value & $\Delta AIC$ & Significance
		}
		\startdata
		2007 Jan. & $6.481^{+0.043}_{-0.031}$ & $0.73^{+0.32}_{-0.29}$ & $235$ & 0.97 (578/595) & $3.21\times10^{-2}$ & $-2.9$ & $\gtrsim2\sigma$ \\
		2008 Jan. & $6.555^{+0.045}_{-0.085}$ & $0.92^{+0.31}_{-0.39}$ & $207$ & 1.04 (618/595) & $1.59\times10^{-2}$ & $-4.3$ & $\gtrsim2\sigma$ \\
		2009 Jan. & $6.604^{+0.060}_{-0.077}$ & $0.59^{+0.38}_{-0.35}$ & $93$ & 1.07 (637/595) & $0.24$ & $1.1$ & $\gtrsim1\sigma$ \\
		2009 Dec. & $6.612^{+0.047}_{-0.045}$ & $1.15^{+0.36}_{-0.48}$ & $169$ & 1.07 (637/595) & $6.10\times10^{-3}$ & $-6.3$ & $\gtrsim2.5\sigma$ \\
		2010 Dec. & $6.585^{+0.030}_{-0.045}$ & $2.01^{+0.49}_{-0.50}$ & $278$ & 1.10 (653/595) & $1.63\times10^{-4}$ & $-13.6$ & $\gtrsim3.5\sigma$ \\
		2011 Dec. & $6.582^{+0.030}_{-0.045}$ & $1.74^{+0.45}_{-0.65}$ & $180$ & 1.05 (625/595) & $1.27\times10^{-3}$ & $-9.4$ & $\gtrsim3\sigma$ \\
		2012 Dec. & $6.699^{+0.035}_{-0.032}$ & $1.66^{+0.58}_{-0.55}$ & $157$ & 1.19 (709/595) & $8.09\times10^{-3}$ & $-5.7$ & $\gtrsim2.5\sigma$ \\
		2014 Nov. & $6.599^{+0.025}_{-0.029}$ & $2.63^{+0.57}_{-0.67}$ & $226$ & 1.09 (646/595) & $2.21\times10^{-5}$  & $-17.6$ & $\gtrsim4\sigma$ \\
		2015 Nov. & $6.708^{+0.031}_{-0.033}$ & $2.64^{+0.55}_{-0.69}$ & $231$ & 1.03 (609/595) & $1.11\times10^{-5}$  & $-19.0$ & $\gtrsim4\sigma$ \\
		2016 Nov. & $6.662^{+0.024}_{-0.026}$ & $3.62^{+0.82}_{-0.78}$ & $263$ & 1.11 (662/595) & $1.18\times10^{-6}$  & $-23.5$ & $\gtrsim4.5\sigma$ \\
		2017 Oct. & $6.671^{+0.040}_{-0.028}$ & $2.92^{+0.76}_{-0.93}$ & $243$ & 1.04 (619/595) & $1.64\times10^{-3}$  & $-8.8$ & $\gtrsim3\sigma$ \\
		2019 Nov. & $6.666^{+0.072}_{-0.052}$ & $1.39^{+1.07}_{-0.99}$ & $99$ & 0.99 (587/595) & $0.37$ & $2.0$ & $\sim1\sigma$ \\
		\enddata
		\tablecomments{Errors are for 1-$\sigma$ confidence range.}
	\end{deluxetable*}

	\clearpage

	\begin{longrotatetable}
		\begin{deluxetable*}{lcccccccccccc}
			\tablecaption{Derived Parameters Based on Spectral Fitting Results\label{tab:derived_par_nH_fixed}}
			\tablenum{6}
			\tabletypesize{\scriptsize}
			\tablehead{
				& 2007 Jan. & 2008 Jan. & 2009 Jan. & 2009 Dec. & 2010 Dec. & 2011 Dec. & 2012 Dec. & 2014 Nov. & 2015 Nov. & 2016 Nov. & 2017 Oct. & 2019 Nov.}
			\startdata
			$R$ &
			$0.764''$ &
			$0.772''$ &
			$0.786''$ &
			$0.787''$ &
			$0.787''$ &
			$0.802''$ &
			$0.805''$ &
			$0.830''$ &
			$0.835''$ &
			$0.842''$ &
			$0.850''$ &
			$0.866''$ \\
			\hline
			\multicolumn{13}{l}{Low-Temperature Plasma} \\
			$n_{\rm e,LT}$ ($f^{-0.5}_{\rm tot}$\,cm$^{-3}$) &
			$1660^{+162}_{-161}$&
			$1789^{+163}_{-130}$&
			$1899^{+153}_{-132}$&
			$2031^{+105}_{-106}$&
			$2057^{+109}_{-92}$&
			$2203^{+112}_{-97}$&
			$2301^{+97}_{-110}$&
			$2240^{+90}_{-97}$&
			$2388^{+106}_{-105}$&
			$2329^{+86}_{-83}$&
			$2579^{+134}_{-117}$&
			$2714^{+195}_{-165}$\\
			$t_{\rm LT}$ ($f^{0.5}_{\rm tot}$\,days) &
			$273^{+28}_{-27}$&
			$257^{+19}_{-23}$&
			$518^{+38}_{-43}$&
			$384^{+23}_{-22}$&
			$460^{+25}_{-26}$&
			$418^{+25}_{-23}$&
			$351^{+17}_{-15}$&
			$422^{+22}_{-20}$&
			$389^{+38}_{-25}$&
			$384^{+20}_{-19}$&
			$328^{+16}_{-18}$&
			$201^{+17}_{-16}$\\
			$t_{\rm u,LT}$ ($f^{0.5}_{\rm tot}$\,days) &
			$1985^{+246}_{-214}$&
			$2109^{+181}_{-222}$&
			$2584^{+303}_{-251}$&
			$2449^{+172}_{-199}$&
			$2970^{+193}_{-243}$&
			$3161^{+146}_{-173}$&
			$3102^{+169}_{-178}$&
			$3791^{+342}_{-229}$&
			$4810^{+232}_{-220}$&
			$3353^{+219}_{-247}$&
			$2867^{+143}_{-171}$&
			$3684^{+251}_{-270}$ \\
			$f_{\rm LT}$ &
			$0.16f$ &
			$0.16f$ &
			$0.19f$ &
			$0.16f$ &
			$0.16f$ &
			$0.13f$ &
			$0.12f$ &
			$0.12f$ &
			$0.10f$ &
			$0.09f$ &
			$0.07f$ &
			$0.04f$ \\
			$M_{\rm LT}$ ($f^{0.5}_{\rm tot}~\Msun$) &
			$0.062$ &
			$0.066$ &
			$0.085$ &
			$0.080$ &
			$0.079$ &
			$0.072$ &
			$0.070$ &
			$0.067$ &
			$0.061$ &
			$0.053$ &
			$0.044$ &
			$0.030$ \\
			\hline
			\multicolumn{13}{l}{High-Temperature Plasma} \\
			$n_{\rm e,HT}$ ($f^{-0.5}_{\rm tot}$\,cm$^{-3}$) &
			$362^{+47}_{-46}$&
			$416^{+45}_{-47}$&
			$448^{+46}_{-45}$&
			$503^{+35}_{-33}$&
			$545^{+35}_{-38}$&
			$573^{+36}_{-37}$&
			$622^{+38}_{-36}$&
			$629\pm35$&
			$646\pm38$&
			$650\pm31$&
			$697^{+46}_{-47}$&
			$675^{+61}_{-62}$ \\
			$t_{\rm HT}$ ($f^{0.5}_{\rm tot}$\,days) &
			$1462^{+239}_{-214}$&
			$1433^{+197}_{-186}$&
			$1375^{+174}_{-161}$&
			$1197\pm103$&
			$1429^{+193}_{-138}$&
			$1232^{+135}_{-99}$&
			$1149\pm100$&
			$1264^{+91}_{-89}$&
			$1053^{+106}_{-79}$&
			$1189^{+85}_{-81}$&
			$1125^{+112}_{-99}$&
			$1072^{+175}_{-133}$ \\
			$t_{\rm u,HT}$ ($f^{0.5}_{\rm tot}$\,days) &
			$3065^{+672}_{-510}$&
			$3555^{+458}_{-460}$&
			$3405^{+483}_{-409}$&
			$2940^{+207}_{-228}$&
			$4650^{+342}_{-337}$&
			$3308^{+217}_{-209}$&
			$3105^{+183}_{-194}$&
			$3989^{+243}_{-233}$&
			$3080^{+183}_{-185}$&
			$3716^{+186}_{-194}$&
			$2854^{+193}_{-189}$&
			$2859^{+268}_{-260}$ \\
			$f_{\rm HT}$  &
			$0.84f$ &
			$0.84f$ &
			$0.81f$ &
			$0.84f$ &
			$0.84f$ &
			$0.87f$ &
			$0.88f$ &
			$0.88f$ &
			$0.90f$ &
			$0.91f$ &
			$0.93f$ &
			$0.96f$ \\
			$M_{\rm HT}$ ($f^{0.5}_{\rm tot}~\Msun$) &
			$0.071$ &
			$0.083$ &
			$0.088$ &
			$0.101$ &
			$0.111$ &
			$0.122$ &
			$0.134$ &
			$0.141$ &
			$0.149$ &
			$0.153$ &
			$0.170$ &
			$0.172$ \\
			\enddata
		\end{deluxetable*}
	\end{longrotatetable}

	\clearpage
	
	\clearpage
	
	\clearpage
	
	\clearpage
	
	\clearpage
	
	\clearpage
	
	\clearpage
	
	\clearpage
	
	\clearpage
	
	\clearpage
	
		\clearpage

	\clearpage
		
\end{document}